\documentclass[twocolumn]{aastex701}
\let\tablenum\relax

\usepackage{url}
\usepackage{threeparttable}
\usepackage{ulem}
\usepackage{natbib}
\usepackage{appendix}
\usepackage{amsmath}
\usepackage{amssymb}
\usepackage{multirow}
\usepackage{float}
\usepackage{physics}
\usepackage{siunitx}
\usepackage{mhchem}
\usepackage{appendix}

\def\ion#1#2{#1$\;${\sc\@roman{#2}}\relax}
\def\lesssim{\mathrel{\hbox{\rlap{\hbox{\lower4pt\hbox{$\sim$}}}\hbox{$<$}}}}
\def\gtrsim{\mathrel{\hbox{\rlap{\hbox{\lower4pt\hbox{$\sim$}}}\hbox{$>$}}}}

\def\red{\textcolor{black}}

\shorttitle{}
\shortauthors{Nakane et al.}
\begin{document}
\title{
Why is GN-z11 Bright, Compact, and Nitrogen Enhanced?\\
Insights from UV Absorption and Emission Diagnostics
%Origin of Compact UV Continuum in GN-z11:\\ Evidence for Massive Stars Shaping Nitrogen Enrichment
}

\author[0009-0000-1999-5472]{Minami Nakane}
\affiliation{Institute for Cosmic Ray Research, The University of Tokyo, 5-1-5 Kashiwanoha, Kashiwa, Chiba 277-8582, Japan}
\affiliation{Department of Physics, Graduate School of Science, The University of Tokyo, 7-3-1 Hongo, Bunkyo, Tokyo 113-0033, Japan}
\email[show]{nakanem@icrr.u-tokyo.ac.jp}  

\author[0000-0002-1049-6658]{Masami Ouchi}
\affiliation{National Astronomical Observatory of Japan, 2-21-1 Osawa, Mitaka, Tokyo 181-8588, Japan}
\affiliation{Institute for Cosmic Ray Research, The University of Tokyo, 5-1-5 Kashiwanoha, Kashiwa, Chiba 277-8582, Japan}
\affiliation{Department of Astronomical Science, SOKENDAI (The Graduate University for Advanced Studies), 2-21-1 Osawa, Mitaka, Tokyo, 181-8588, Japan}
\affiliation{Kavli Institute for the Physics and Mathematics of the Universe (WPI), The University of Tokyo, 5-1-5 Kashiwanoha, Kashiwa, Chiba 277-8583, Japan}
\email{ouchims@icrr.u-tokyo.ac.jp}



\begin{abstract}

%250 words
We investigate the UV spectrum of GN-z11, a luminous, compact galaxy with strong nitrogen lines, at $z=10.60$, using deep JWST/NIRSpec high-resolution IFU and medium-resolution MSA spectra assembled from the JADES, SPURS, and GO programs. After optimized reduction and extraction of the IFU data including an evaluation of statistical and systematic uncertainties, we obtain mutually consistent spectra from the high- and medium-resolution observations. After carefully accounting for the data quality limitations, we identify prominent P-Cygni profiles in N\textsc{v}$\lambda\lambda1238,1243$, Si\textsc{iv}$\lambda\lambda1394,1403$, and C\textsc{iv}$\lambda\lambda1548,1550$, together with broad N\textsc{iv}]$\lambda\lambda1483,1486$ emission (FWHM $\sim1600$ km s$^{-1}$). The P-Cygni profiles resemble those of massive stars such as O-type stars and luminous blue variables (LBVs), while the broad N\textsc{iv}] emission resembles that of nitrogen-sequence Wolf-Rayet (WN) stars. We fit stellar and active galactic nuclei (AGN) UV spectral models and find that the stellar models are strongly preferred over the AGN models ($\Delta$WAIC $=\red{-20}$), with the preference driven primarily by the NV P-Cygni profile. These results indicate that the luminous, compact UV continuum of GN-z11 is dominated by massive stars. We derive electron densities from C\textsc{iii}]$\lambda\lambda1907,1909$, N\textsc{iii}]$\lambda\lambda1747-1754$, and N\textsc{iv}], with the nitrogen diagnostics extending well beyond the C\textsc{iii}]-based limit and reaching densities of $\gtrsim10^{6.5}$ cm$^{-3}$ for N\textsc{iv}], indicating physically distinct carbon- and nitrogen-emitting nebular components. These findings suggest that the apparent nitrogen enhancement inferred for GN-z11 as a whole may arise when strong narrow nitrogen emission originates from dense gas locally enriched in nitrogen by WN stellar winds and photoionized by nearby massive stars within the same star-forming region.

\end{abstract}


%% subsec tion for cross-referencing in a \ref command.

\section{Introduction}
\label{sec:introduction}

After the launch of the James Webb Space Telescope (JWST), an increasing number of galaxies at $z\gtrsim10$ have been spectroscopically confirmed (e.g., \citealt{Haro2023a,Bunker2023,Curtis-Lake2023,Castellano2024,Fujimoto2024,Naidu2025,Napolitano2025,Donnan2026}). One of the major discoveries is the unexpectedly large population of UV-luminous galaxies at $z\gtrsim10$ compared to theoretical model predictions (e.g., \citealt{Finkelstein2023,Harikane2024,Harikane2025}). Proposed explanations include high star formation efficiency (e.g., \citealt{Fukushima2021,Dekel2023,Yung2024}), active galactic nucleus (AGN) activity (e.g., \citealt{Harikane2023,Hegde2024}), radiation-driven outflows (e.g., \citealt{Ferrara2023,Ferrara2024}), bursty star formation (e.g., \citealt{Munoz2023,Sun2023,Chen2026a,Harikane2026}), and a top-heavy initial mass function (IMF; e.g., \citealt{Omukai2005,Chon2021}), but the dominant physical mechanism remains uncertain. Understanding the origin of these luminous galaxies is therefore a key question in studies of galaxy formation during the epoch of reionization.

A representative example is GN-z11 at $z=10.60$ with a UV magnitude of $M_\mathrm{UV}=-21.5$, one of the brightest galaxies known at $z>10$ \citep{Oesch2016,Bunker2023}. JWST/NIRCam imaging reveals an extremely compact UV morphology with an effective radius of only $r_\mathrm{eff}=64$ pc \citep{Tacchella2023} while JWST/NIRSpec spectroscopy shows strong high-ionization UV emission lines, including N \textsc{iv}] $\lambda\lambda1483,1486$ \citep{Bunker2023,Maiolino2024}. Previous studies have suggested unusually high N/O and N/C abundance ratios in the ionized gas (e.g., \citealt{Bunker2023,Cameron2023,Isobe2023b,Senchyna2024}). The electron densities inferred from the UV emission lines reach $n_\mathrm{e}>10^9$ cm$^{-3}$, comparable to those found in AGN broad-line regions (BLRs; \citealt{Maiolino2024}). However, JWST/MIRI medium-resolution (MRS) spectroscopy does not reveal a broad H$\alpha$ component expected from a classical BLR \citep{Alvarez-Marquez2025}. MIRI imaging indicates that the flux excess in the optical red continuum is inconsistent with hot dust emission from an unobscured AGN \citep{Crespo-Gomez_2026}. These apparently contradictory observational results leave the origin of the compact UV emission, the ionizing source, and the nitrogen-enhanced gas in GN-z11 unresolved.

\red{Extremely high N/O ratios are not unique to GN-z11, but have also been identified in a growing number of high-redshift galaxies (e.g., \citealt{Isobe2023b,Larson2023,Ubler2023,Castellano2024,Ji2024,Schaerer2024,Topping2024,Navarro-Carrera2025,Topping2025a,Zhang2026}) and rare low-redshift metal-poor galaxies (e.g., \citealt{Bhattacharya2025}). These nitrogen-enhanced galaxies show N/O ratios of $\mathrm{[N/O]}\gtrsim0$, substantially higher than those of typical local galaxies (e.g., \citealt{Izotov2006,Berg2021}) and Galactic H \textsc{ii} regions \citep{Garcia-Rojas2007} at similar metallicities, but comparable to those of some stars in local globular clusters (GCs; \citealt{Carretta2005}), motivating a possible connection with GC formation (e.g., \citealt{Isobe2023b,Senchyna2024,Ji2026}). While asymptotic giant branch (AGB) stars can contribute substantially to nitrogen enrichment on relatively long timescales, the young ages of metal-poor nitrogen-enhanced galaxies at high redshift motivate more rapid enrichment channels, including Wolf-Rayet (WR) stars (e.g., \citealt{Cameron2023,Fukushima2024,Kobayashi&Ferrara2024,Watanabe2024,Watanabe2026}), supermassive stars (SMSs; e.g., \citealt{Charbonnel2023,Nagele2023,Nandal2025,Ebihara2026,Watanabe2026}), and tidal disruption events (TDEs; e.g., \citealt{Cameron2023,Watanabe2026}). However, subsequent core-collapse supernovae (CCSNe) eject substantial amounts of oxygen and can rapidly lower the enhanced N/O ratio. Several scenarios have therefore been proposed to produce and maintain high N/O ratios, including dual starbursts \citep{Kobayashi2024}, chemically differential outflows \citep{Rizzuti2025}, and failed supernovae/direct collapse \citep{Watanabe2024,Watanabe2026}. The origin of the extreme nitrogen enhancement remains under debate.}

In this study, we analyze UV spectra of GN-z11 at $z=10.60$ using deep JWST/NIRSpec medium-resolution MSA and high-resolution IFU data \red{taken by JADES \citep{Bunker2023,Eisenstein2023}, SPURS \citep{Chen2026a}, and GO-5086 \citep{Maiolino2026}}. \red{We note that an independent analysis of the medium-resolution spectrum of GN-z11 by the SPURS team is presented in \citet{Chen2026b}.} This paper is organized as follows. In Section \ref{sec:data}, we describe the data used in this study and compare the MSA and IFU spectra. Section \ref{sec:origin} investigates the origin of the UV spectral features and continuum. In Section \ref{sec:gas_nitrogen}, we present the electron density measurements and discuss their implications for nitrogen enhancement. Section \ref{sec:summary} summarizes our major findings. We assume a standard $\Lambda$CDM cosmology with $\Omega_\Lambda=0.7$, $\Omega_m=0.3$, and $H_0=70$ km s$^{-1}$ Mpc$^{-1}$. All magnitudes are in the AB system \citep{Oke&Gunn1983}. \red{Throughout this paper, we utilize the solar abundance ratios of $12+\log(\mathrm{O/H})_\odot=8.69$, $\log(\mathrm{C/O})_\odot=-0.26$, and $\log(\mathrm{N/O})_\odot=-0.83$ \citep{Asplund2009}. The notation $[X/Y]$ is defined as $\log(X/Y)$ relative to the solar abundance ratio, i.e., $[X/Y]=\log(X/Y)-\log(X/Y)_\odot$.}

\begin{figure*}
    \centering
    \includegraphics[width=0.99\linewidth]{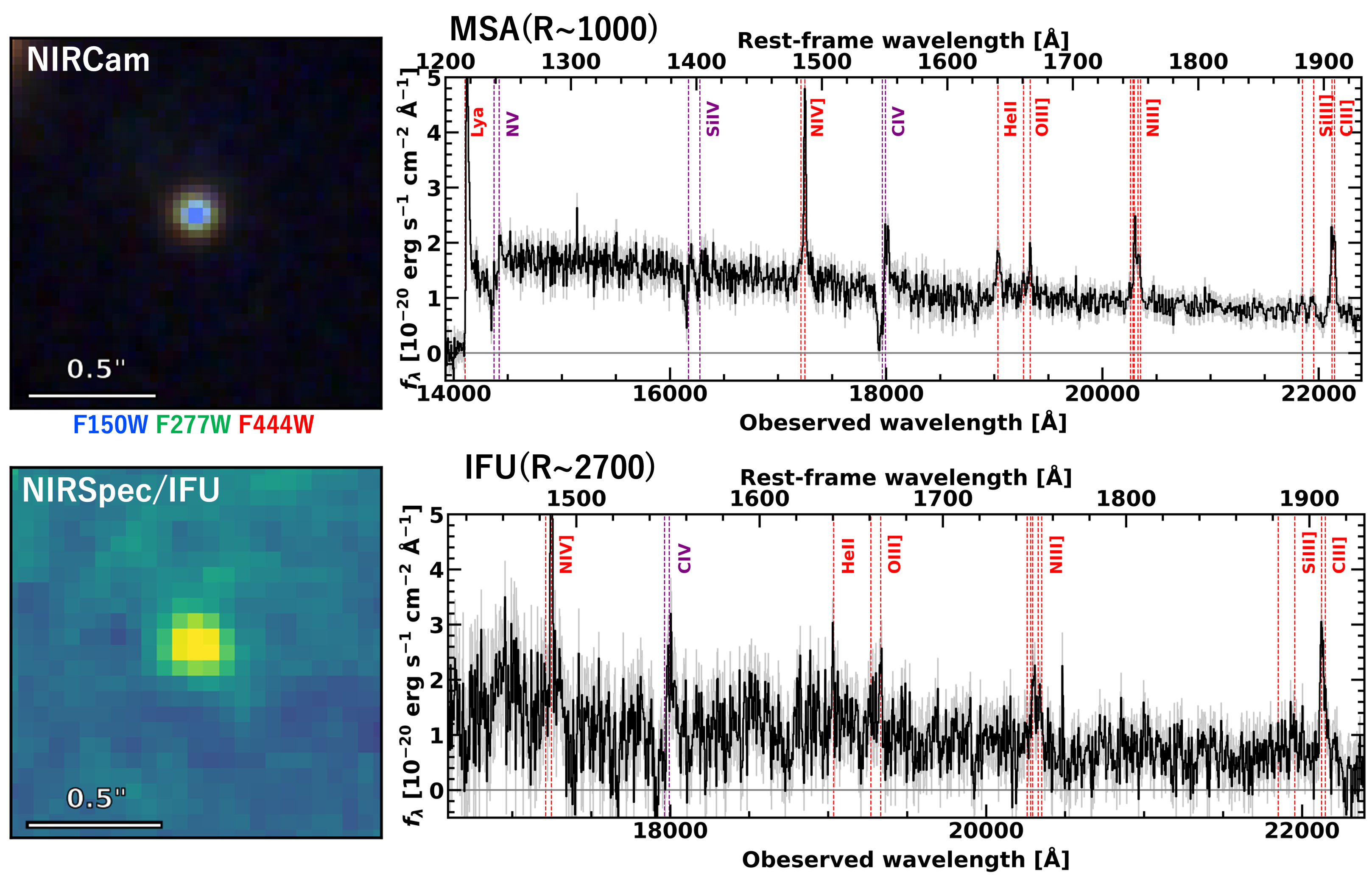}
    \caption{Top: NIRCam images \red{from JADES} (R: F444W, G: F277W, B: F150W; left) and NIRSpec/MSA medium-resolution spectra \red{from SPURS and JADES} (right). The red and purple dashed lines indicate the emission lines and P-Cygni lines, respectively. Bottom: NIRSpec/IFU G235H image collapsed around the N \textsc{iv}] $\lambda\lambda1483,1486$ emission lines (left) and NIRSpec/IFU G235H spectrum (right), \red{both from GO-5086}.}
    \label{fig:data}
\end{figure*}

\section{Data} 
\label{sec:data}

\subsection{NIRSpec/MSA}
\label{subsec:msa}
We use the NIRSpec/MSA spectra of GN-z11 observed as part of the Spectroscopic Ultra-deep Reionization-era Survey (SPURS; GO-9214, PIs: C. Mason \& D. Stark; \citealt{Chen2026a}) and JWST Advanced Deep Extragalactic Survey (JADES; GTO-1181, PI: D. Eisenstein; \citealt{Eisenstein2023}). The SPURS (JADES) NIRSpec/MSA observations were conducted with the medium-resolution ($R\sim1000$) grating/filter pairs of G140M/F100LP (G140M/F070LP), G235M/F170LP, and G395M/F290LP, which cover $1.0-1.6$ ($0.7-1.3$), $1.7-3.1$, and $2.9-5.1$ $\mu$m, respectively. The exposure times of the SPURS observations for the G140M, G235M, and G395M gratings are $29.2$, $7.9$, and $2.9$ hours, respectively. \red{The exposure time of the JADES observations for GN-z11 is $6.2$ hours for each grating. In total, the exposure times of GN-z11 are $35.4$, $14.1$, and $9.1$ hours for G140M, G235M, and G395M, respectively.} We reduce the SPURS and JADES data obtained from the Mikulski Archive for Space Telescopes (MAST) portal, using \texttt{msaexp} (v0.9.18; \citealt{Brammer2023}) based on the JWST pipeline (ver.1.16.0) with the Calibration Reference Data System (CRDS) \texttt{jwst\_1322.pmap} calibration file, following the previous studies \citep{de_Graaff2024,Heintz2025,Valentino2025}. To achieve higher signal-to-noise (S/N) ratios, we co-add the SPURS and JADES spectra for each grating. We first multiply each spectrum by a scale factor. We determine the scale factor by comparing the JWST/NIRCam photometry and filter-convolved NIRSpec spectra at $1.4-3.2$ $\mu$m in the observed frame to focus on the rest-frame UV spectra and avoid the Lyman break. For the NIRCam photometry, we use \red{the $r=0\farcs1$ aperture photometry from} the JADES Data Release 5 catalog \citep{Eisenstein2026,Robertson2026}. We then derive the co-added spectra and their associated $1\sigma$ uncertainties by calculating inverse-variance-weighted mean fluxes from the scaled spectra. The co-added MSA spectra are presented in the top panel of Figure \ref{fig:data}.

\begin{figure*}[ht!]
    \centering
    \includegraphics[width=0.99\linewidth]{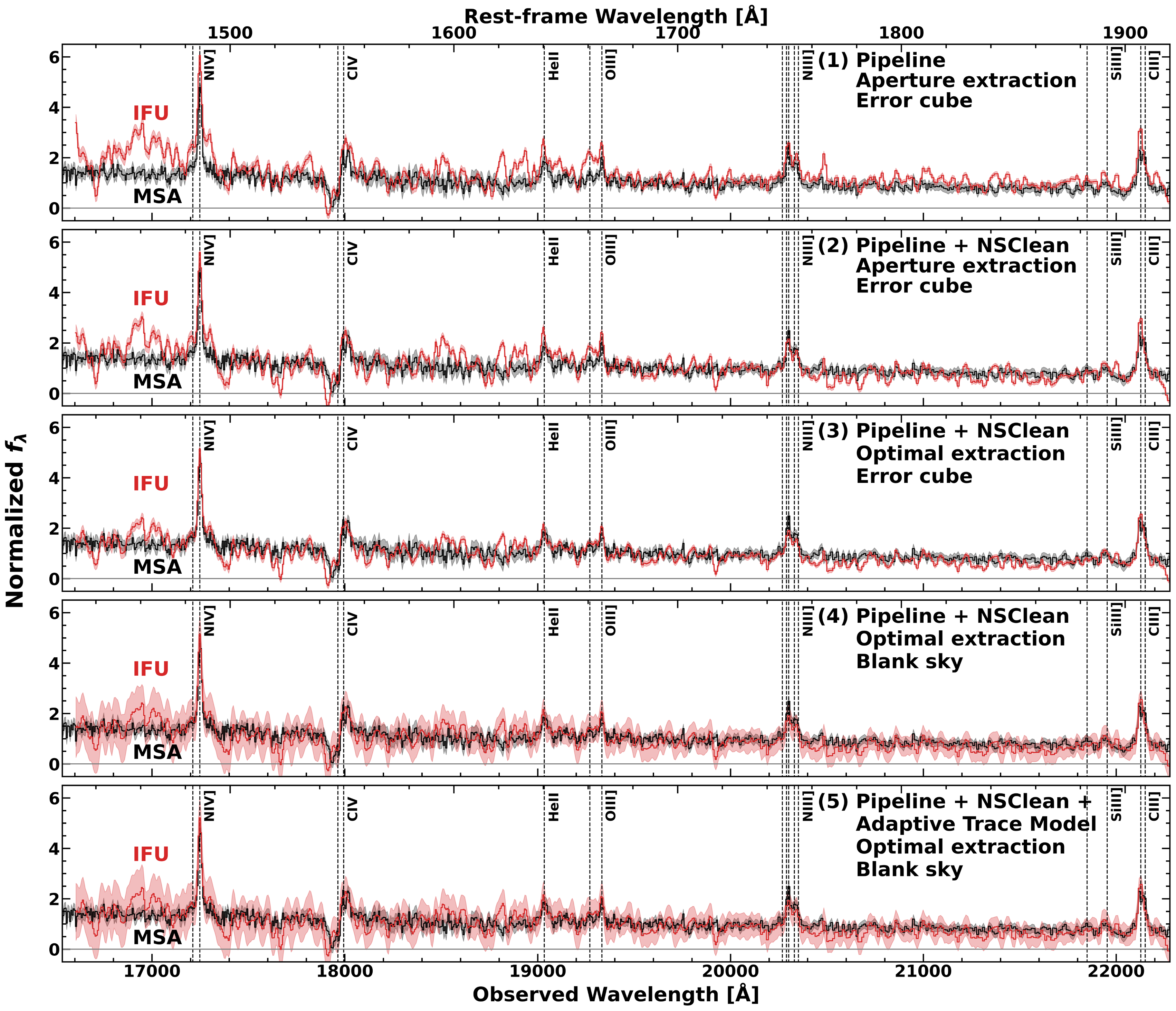}
    \caption{Comparison between the MSA and IFU spectra \red{obtained through SPURS/JADES and GO-5086, respectively}. From top to bottom, we compare the resolution-matched MSA and IFU spectra with different reduction, extraction, and error spectrum estimation (see Section \ref{subsec:comparison}). The black (red) solid line and shaded region show the MSA (IFU) spectrum and its $1\sigma$ uncertainty, respectively. The black dashed lines indicate the emission and P-Cygni lines detected in the spectra (see Section \ref{subsec:UV_feature}).}
    \label{fig:spec_comparison}
\end{figure*}

\subsection{NIRSpec/IFU}
\label{subsec:ifu}

We utilize the NIRSpec/IFU data of GN-z11 that were obtained through a GO program (GO-5086; PI: R. Maiolino; \citealt{Maiolino2026}). The observations were carried out with the high-resolution ($R\sim2700$) grating/filter pair of G235H/F170LP, covering $1.7-3.1$ $\mu$m. \red{Out of the $39$ dithers, we discard one exposure because the guiding camera locked onto a hot pixel, as described in \citet{Maiolino2026}. After removing this frame, the total exposure time is $38.8$ hours.} We reduce the IFU data obtained from the MAST portal, using the JWST Science Calibration Pipeline (version 2.0.1; \citealt{Bushouse2026}) with the CRDS calibration file \texttt{jwst\_1535.pmap}. Our reduction follows the procedures presented by \citet{Kiyota2025}, which are built on the publicly available script \citep{Rigby2024} developed by the TEMPLATES team that is the JWST Early Release Science (ERS) program (ERS-1355; PI: Jane R. Rigby; \citealt{Rigby2025}). We first process each exposure with \texttt{calwebb\_detector1} as stage 1, which performs detector-level corrections. Before running \texttt{calwebb\_spec2}, we correct the $1/f$ noise of the data using \texttt{NSClean} \citep{Rauscher2024}, which is not included in the default pipeline. We then process each stage-1 product with \texttt{calwebb\_spec2} as stage 2, including world coordinate system and wavelength corrections, flat-fielding, path-loss corrections, and flux calibration. We conduct drizzle weighting, resulting in the spaxel scale of $0\farcs06$. Instead of using stage 3, we combine stage 2 with \texttt{reproject}(version 0.13.1; \citealt{Robitaille2024}) and \texttt{reproject\_interp} package through median stacking, which effectively reduces outliers. Finally, we estimate and subtract the background independently for each wavelength channel to obtain the final background-subtracted 3D data cube. \red{For comparison, we also construct an alternative IFU cube using \texttt{adaptive\_trace\_model} method \citep{Law2026}, designed to mitigate resampling artifacts arising from spatial undersampling. We apply the \texttt{adaptive\_trace\_model} method during \texttt{calwebb\_spec2} to the stage-1 products, using the JWST pipeline ver. 3.0.0 \citep{Bushouse2026b} and \texttt{jwst\_1584.pmap}.}

The 1D spectrum is extracted using the optimal extraction method \citep{Horne1986}, which has been applied to NIRSpec/IFU data and provides a higher spectral S/N ratio than aperture extraction \citep{D'Eugenio2026}. In the optimal extraction, the extracted flux is calculated as the inverse-variance weighted sum of the flux in each spaxel using a model of the object's spatial profile. We adopt the S\'ersic profile derived from the NIRCam/F200W image by \citet{Nakane2026} to describe the morphology of GN-z11. The corresponding weight map is generated by convolving the S\'ersic model by the IFU point-spread function (PSF), which is modeled using \texttt{STPSF} \citep{Perrin2025}. The resulting 1D spectrum is shown in the bottom panel of Figure \ref{fig:data}.

\begin{figure*}
    \centering
    \includegraphics[width=0.85\linewidth]{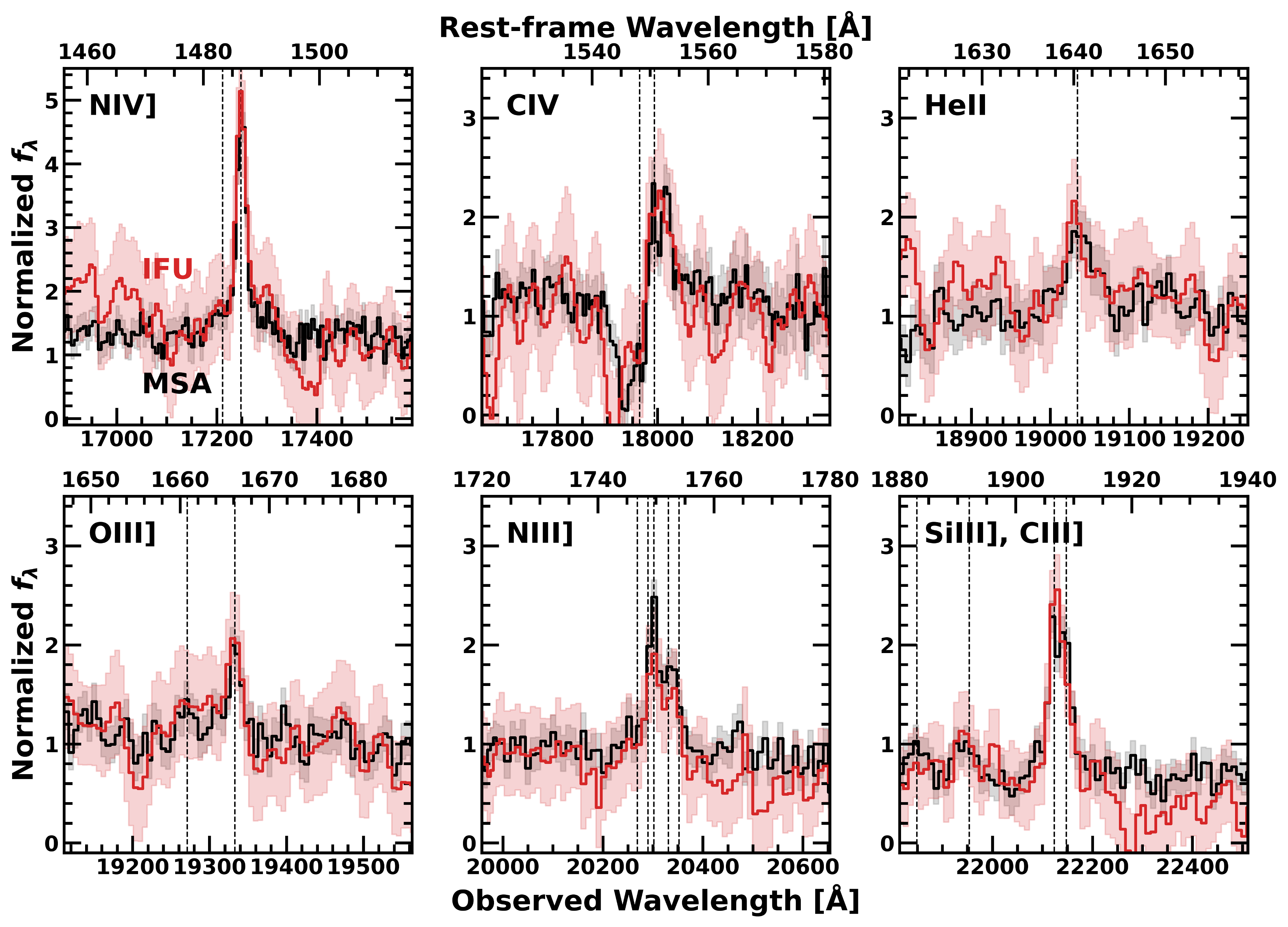}
    \caption{Comparison between the MSA and IFU spectra \red{from SPURS/JADES and GO-5086, respectively,} for the N \textsc{iv}], C \textsc{iv}, He \textsc{ii}, O \textsc{iii}], N \textsc{iii}], Si \textsc{iii}],  and C \textsc{iii}] lines. The lines and shaded regions represent the same as in Figure \ref{fig:spec_comparison}.} 
    \label{fig:line_comparison}
\end{figure*}

\subsection{Comparison of MSA and IFU spectra}
\label{subsec:comparison}

For NIRSpec/IFU observations, previous studies have shown that the resulting spectra can be sensitive to the data reduction and extraction procedures (e.g., \citealt{Perna2023,Rauscher2024,D'Eugenio2026}). In this section, we investigate how different reduction and extraction methods affect the extracted IFU spectrum. In Figure \ref{fig:spec_comparison}, we compare the MSA and IFU spectra after smoothing and resampling the IFU spectra to match the spectral resolution of $R\sim1000$. We compare \red{five} versions of the IFU spectrum: (1) a spectrum reduced with the default pipeline and extracted using a circular aperture with a diameter of $d=0\farcs3$, with the associated error spectrum taken from the pipeline-generated error cube; (2) the same as (1), but with the \texttt{NSClean} correction applied before \texttt{calwebb\_spec2}; (3) the same as (2), but extracted using the optimal extraction method; (4) the same as (3) but with the error spectrum estimated from randomly selected blank-sky regions instead of the pipeline error cube\red{; (5) the same as (4) but with the \texttt{adaptive\_trace\_model} method applied while running \texttt{calwebb\_spec2}.}

Figure \ref{fig:spec_comparison} shows that spectrum (1) exhibits significant continuum fluctuations relative to the MSA spectra particularly around $\sim1.7$ and $\sim1.9\ \mu$m. The \texttt{NSClean} correction partially suppresses these fluctuations, as shown in spectrum (2), while the optimal extraction further improves the continuum quality in spectrum (3). However, the error spectrum derived from the pipeline error cube is significantly underestimated because it does not fully account for correlated noise, which may lead to spurious emission- or absorption-line detections. In contrast, the error spectrum estimated from randomly selected blank-sky regions naturally accounts for these correlations, resulting in spectrum (4), which is in good agreement with the MSA spectra. \red{We further apply the \texttt{adaptive\_trace\_model} method described above and show the resulting spectrum as spectrum (5). Both the extracted spectrum and its uncertainties remain very similar to those of spectrum (4), likely because our optimal extraction combines multiple spatial pixels and reduces the impact of resampling artifacts at the individual-pixel level. Since the adaptive trace model correction does not significantly alter the resulting spectrum or its uncertainties in our analysis, we adopt spectrum (4), which combines the \texttt{NSClean} correction, optimal extraction, and blank-sky-based error estimation, throughout the remainder of this paper as our fiducial IFU spectrum.} Figure \ref{fig:line_comparison} presents zoomed-in views of the UV emission lines covered by both the co-added MSA and adopted IFU spectra, demonstrating the consistency of the line profiles.

\section{Origin of Compact UV Continuum}
\label{sec:origin}

\subsection{UV Spectral Features}
\label{subsec:UV_feature}

As shown in Figure \ref{fig:data}, we identify various line features in the UV spectra of GN-z11, which have also been reported in the literature (e.g., \citealt{Bunker2023,Maiolino2024,Senchyna2024,Isobe2026,Nakane2026,Chen2026b}): emission lines of Ly$\alpha$, N \textsc{iv}] $\lambda\lambda1483,1486$, C \textsc{iv} $\lambda\lambda1548,1550$, He \textsc{ii} $\lambda1640$, O \textsc{iii}] $\lambda\lambda1661,1666$, N \textsc{iii}] $\lambda\lambda1747-1754$ (quintet), Si \textsc{iii} $\lambda\lambda1883,1892$, and C \textsc{iii}] $\lambda\lambda1907,1909$; P-Cygni signatures of N \textsc{v} $\lambda\lambda1238,1243$, Si \textsc{iv} $\lambda\lambda1394,1403$, and C \textsc{iv} $\lambda\lambda1548,1550$ lines. These spectral features arise from a wide range of physical processes. The resonance lines exhibiting P-Cygni profiles have frequently been associated with stellar winds from massive stars. In contrast, the semi-forbidden emission lines are primarily produced in photoionized gas, and their strengths depend on the ionization state, density, and chemical abundances of the emitting regions. The coexistence of broad P-Cygni signatures and narrow nebular emission therefore indicates that multiple physical components contribute to the UV spectrum of GN-z11. \red{In addition, as shown in Figure \ref{fig:line_comparison}, the N \textsc{iv}] profile exhibits excess emission in the wings relative to the other narrow nebular lines in both the MSA and IFU spectra, motivating the decomposition of the narrow and broad components described below.}

We conduct emission-line fitting with Gaussian profiles to the N \textsc{iv}], He \textsc{ii}, O \textsc{iii}], N \textsc{iii}], Si \textsc{iii}], and C \textsc{iii}] lines, together with a continuum approximated by a linear function. Since Ly$\alpha$, N \textsc{v}, Si \textsc{iv}, and C \textsc{iv} are analyzed with spectral models in Section \ref{subsec:fitting}, we do not fit these lines independently with Gaussian profiles here. We conduct single-component fitting to each transition, where we simultaneously fit multiple transitions from the same ion at nearby wavelengths (e.g., N \textsc{iv}] $\lambda\lambda1483,1486$) with common line widths. We fix the line centroid according to the systemic redshift $z_\mathrm{sys}=10.6034$, which is estimated from multiple emission lines in \citet{Bunker2023}. For the N \textsc{iii}] multiplet, the line ratios of the N \textsc{iii}] $\lambda\lambda1747,1752$ and N \textsc{iii}] $\lambda\lambda1749,1754$ pairs are fixed, following the approach of \citet{Maiolino2024}, because each pair arises from the same upper level. We adopt theoretical ratios of $F_{1747}/F_{1752}=0.17$ and $F_{1754}/F_{1749}=1.07$ calculated with \texttt{PyNeb} \citep{Luridiana2015}. We derive an intrinsic line width $\sigma_\mathrm{int}$ with $\sigma_\mathrm{int}=\sqrt{\sigma_\mathrm{obs}^2-\sigma_\mathrm{inst}^2}$, where $\sigma_\mathrm{obs}$ is an observed line width and $\sigma_\mathrm{inst}$ is a line width of a line-spread function (LSF) approximated by a Gaussian function. We adopt the LSFs derived by \citet{Isobe2023a}, scaling its width by a factor of $0.5$ because the spectral resolution is approximately twice the nominal resolution for compact sources \citep{de_Graaff2024}. We perform Monte Carlo Markov Chain (MCMC) fitting using \texttt{emcee} \citep{Foreman2013}, assuming flat priors for the free parameters of line fluxes, line widths, slope and intercept of the linear function. We derive the best-fit parameter and its $1\sigma$ uncertainty from the mode (i.e., the peak of the posterior distribution) and the 68\% highest posterior density interval (HPDI; i.e., the narrowest interval containing 68\%), respectively. \red{For N \textsc{iv}], we additionally perform two-component fitting by adding a single broad component with the line flux and width as free parameters and the centroid fixed to that of the narrow N \textsc{iv}] $\lambda1486$ component.} We compare the single-component and two-component fitting to N \textsc{iv}] by evaluating the goodness of the fits with the widely applicable information criterion (WAIC; \citealt{Watanabe2010}), which is suitable for Bayesian parameter estimation.

The best-fit line properties are summarized in Table \ref{tab:gaussian}. The He \textsc{ii}, O \textsc{iii}], N \textsc{iii}], and C \textsc{iii}] lines are detected at high S/N ratios of $6.2$, $7.1$, $13.1$, and $12.8$, respectively. The Si \textsc{iii}] lines are tentatively detected at $\mathrm{S/N}=3.1$ while the S/N ratios of individual flux measurements are marginal. For \red{N \textsc{iv}]}, the difference in the WAIC between the models with and without a broad component is $\Delta\mathrm{WAIC}=-9.7$ and the S/N ratio of the broad component flux is $3.7$, indicating that the model including an N \textsc{iv}] broad component is statistically preferred. We note that when adopting the physically-motivated priors that connect the narrow doublet flux ratios with the electron density, the significance of the broad component increases to $\mathrm{S/N}=7.1$ because the weak N \textsc{iv}] $\lambda1483$ line \red{has a smaller effect on the fit}  (see Section \ref{subsubsec:ne_measurement}). The line widths of the narrow components are overall in the range of $\sim300-400$ km s$^{-1}$, slightly larger than the measurements of [O \textsc{iii}] $\lambda5007$ ($\mathrm{FWHM}=189\pm25$ km s$^{-1}$) and H$\alpha$ ($\mathrm{FWHM}=231\pm52$ km s$^{-1}$) obtained from MIRI/MRS \citep{Alvarez-Marquez2025}. The He \textsc{ii} line shows a tentative broad component with $\mathrm{FWHM}=752_{-144}^{+144}$ km s$^{-1}$ compared with the narrow nebular lines, although the uncertainty of the width measurement remains relatively large. Noticeably, the preferred N \textsc{iv}] broad component has $\mathrm{FWHM}=1638_{-412}^{+495}$ km s$^{-1}$, significantly broader than the narrow components. These broad emission components may arise from a distinct physical component. The origin of the broad N \textsc{iv}] component and the possible broad He \textsc{ii} emission and their implications are further discussed in Section \ref{subsec:wind_outflow}.

\begin{table}
    \caption{Line Properties from Gaussian Fitting.}
    \begin{tabular}{cccc}
    \hline
    \hline
    Line & & Flux & FWHM \\
     & & [$10^{-19}$ erg s$^{-1}$ cm$^{-2}$] & [km s$^{-1}$] \\
     (1) & & (2) & (3) \\
    \hline
    N \textsc{iv}] & total & $11.7_{-1.2}^{+0.5}$ & - \\
     & $\lambda1483$ & $<1.4^\mathrm{a}$ & $336_{-30}^{+23}$ \\
     & $\lambda1486$ & $7.1_{-1.0}^{+0.2}$ & $336_{-30}^{+23}$ \\
     & broad & $4.4_{-1.5}^{+0.9}$ & $1638_{-412}^{+495}$ \\
    He \textsc{ii} & $\lambda1640$ & $3.8_{-0.6}^{+0.6}$ & $752_{-144}^{+144}$ \\ 
    O \textsc{iii}] & total & $3.7_{-0.7}^{+0.4}$ & - \\
     & $\lambda1661$ & $1.2_{-0.5}^{+0.2}$ & $396_{-72}^{+96}$ \\
     & $\lambda1666$ & $2.5_{-0.5}^{+0.3}$ & $396_{-72}^{+96}$ \\    
    N \textsc{iii}] & total & $7.5_{-0.7}^{+0.5}$ & - \\
     & $\lambda1747$ & $0.4_{-0.1}^{+0.1}$ & $288_{-51}^{+51}$ \\
     & $\lambda1749$ & $0.9_{-0.3}^{+0.2}$ & $288_{-51}^{+51}$ \\
     & $\lambda1750$ & $3.1_{-0.4}^{+0.3}$ & $288_{-51}^{+51}$ \\
     & $\lambda1752$ & $2.1_{-0.3}^{+0.3}$ & $288_{-51}^{+51}$ \\
     & $\lambda1754$ & $1.0_{-0.3}^{+0.2}$ & $288_{-51}^{+51}$ \\
    Si \textsc{iii}] & total & $1.8_{-0.7}^{+0.4}$ & - \\
    & $\lambda1883$ & $0.8_{-0.4}^{+0.3}$ & $462_{-119}^{+279}$ \\
    & $\lambda1892$ & $1.1_{-0.8}^{+0.1}$ & $462_{-119}^{+279}$ \\
    C \textsc{iii}] & total & $7.3_{-0.4}^{+0.7}$ & - \\
    & $\lambda1907$ & $3.8_{-0.3}^{+0.7}$ & $367_{-63}^{+146}$ \\
    & $\lambda1909$ & $3.3_{-0.4}^{+0.4}$ & $367_{-63}^{+146}$ \\    
    \hline    
    \end{tabular}
    \par
    \vspace{0.02\hsize}
    \footnotesize{(1): Line name. (2): Flux. (3): Line width.\\
    a: $3\sigma$ upper limit for a posterior distribution peaking at zero flux.}
    \label{tab:gaussian}
\end{table}

\subsection{Comparison with Massive Stars}
\label{subsec:massive_stars}

\begin{figure*}
    \centering
    \includegraphics[width=0.99\linewidth]{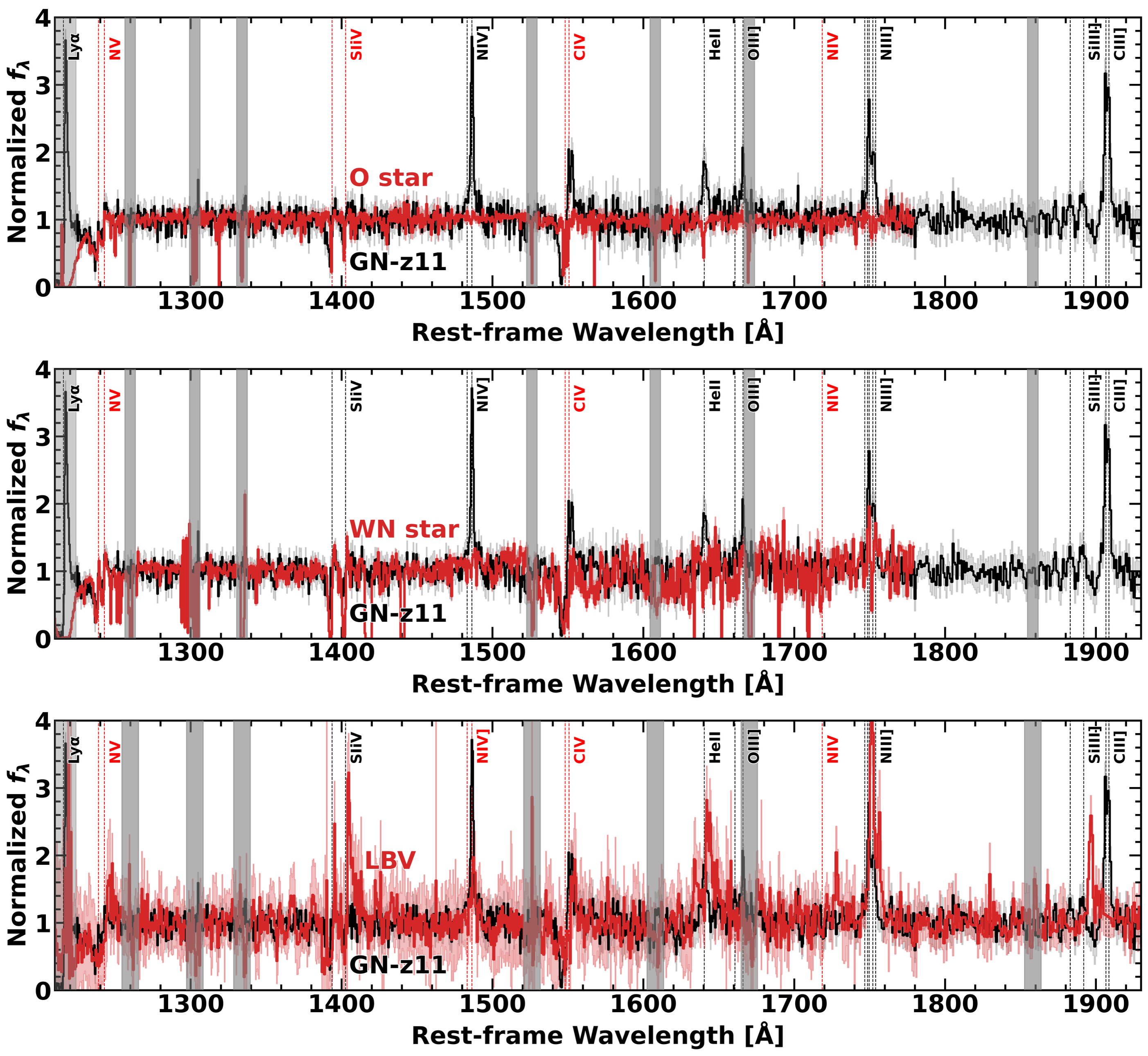}
    \caption{Spectral comparison between GN-z11 and massive stars. The black solid lines and light-gray shaded regions indicate the MSA spectra of GN-z11 and the associated $1\sigma$ uncertainties, respectively. The red solid lines and shaded regions show the spectra of massive stars (top: O star, middle: WN star, bottom: LBV) and their $1\sigma$ uncertainties, respectively. The red labels and dashed lines present line features that are comparable to those of massive stars and their rest-frame wavelengths. The black labels and dashed lines denote the same as red ones but for the line features that appear to be difficult to explain by the massive stars.}
    \label{fig:ullyses_lbv}
\end{figure*}

The rest-frame UV spectra of GN-z11 exhibit prominent P-Cygni profiles, which are commonly interpreted as signatures of stellar winds from massive stars (e.g., \citealt{Pettini2002,Leitherer2011,Steidel2016,Chisholm2019}). To investigate the stellar origin of these features, Figure \ref{fig:ullyses_lbv} compares the UV spectrum of GN-z11 with massive stars in the local Universe, including an O star, a WN star, and a luminous blue variable (LBV). As representative O- and WN-type stars, we adopt AV-446 (O 6.5V) and SK-69D175 (WN 11h), respectively, from the Hubble UV Legacy Library of Young Stars as Essential Standards (ULLYSES) program \citep{Roman-Duval2025}. These stars were selected because they provide the closest empirical match to the observed P-Cygni profiles among the currently available ULLYSES spectra, allowing a qualitative comparison of the stellar wind features. For the LBV, we use HD5980 during its 1994 eruption, observed with the International Ultraviolet Explorer (IUE), whose spectrum is characterized by strong N \textsc{iii}] and N \textsc{iv}] emission lines \citep{Georgiev2011,Hillier2019}. When comparing observed UV spectra, it is important to consider possible contamination from interstellar (ISM) and circumgalactic (CGM) absorption, which can change the intrinsic stellar P-Cygni profiles. In particular, the Si \textsc{iv} and C \textsc{iv} lines are frequently affected by absorption associated with the ISM and CGM in both local massive stars and star-forming galaxies. By contrast, such contamination is expected to be much less significant for N \textsc{v} because its high ionization potential (77.5 eV) makes it difficult to produce in typical ISM/CGM gas. We therefore regard the N \textsc{v} profiles as the most reliable feature for comparison with local massive stars. The N \textsc{v} profiles of GN-z11 closely resemble those of the O star, WN star, and LBV. Although the Si \textsc{iv} and C \textsc{iv} profiles may be partly affected by interstellar absorption, their overall profiles are also broadly consistent with those of the comparison stars except for Si \textsc{iv} of the WN star and LBV. A notable characteristic of GN-z11 is the absence of strong N \textsc{iv} $\lambda1718$ and He \textsc{ii} emission typically seen in early-type WN stars. However, the representative stars adopted here, AV-446 (O 6.5V), SK-69D175 (WN 11h), and HD5980 during its eruptive phase (LBV), also exhibit weak or absent N \textsc{iv} emission. The lack of prominent N \textsc{iv} emission in GN-z11 is consistent with these comparison spectra. In contrast, the LBV exhibits prominent N \textsc{iv}], He \textsc{ii}, and N \textsc{iii}] emission in addition to its P-Cygni profiles. The similarity of these nitrogen emission lines to those observed in GN-z11 makes the LBV spectrum a useful empirical comparison, while the physical origin of these emission features is investigated in the following sections. Overall, the UV spectrum of GN-z11 shares many of its spectral features with those of local massive stars. In the following section, we investigate whether these similarities are supported by quantitative spectral fitting.

\subsection{Spectral Fitting}
\label{subsec:fitting}

In Section \ref{subsec:massive_stars}, we identify several similarities between the rest-frame UV spectra of GN-z11 and those of local massive stars. However, such empirical similarities alone do not uniquely establish the origin of the UV continuum. In particular, P-Cygni-like profiles can also be reproduced by the combination of nebular emission and blue-shifted ISM/CGM absorption. Moreover, even the N \textsc{v} profile alone does not uniquely distinguish between stellar and AGN ionizing sources. We therefore conduct UV spectral fitting to evaluate whether the observed spectrum is better reproduced by stellar or AGN models, focusing on the N \textsc{v} $\lambda\lambda1238,1243$, Si \textsc{iv} $\lambda\lambda1394,1403$, and C \textsc{iv} $\lambda\lambda1548,1550$ P-Cygni signatures. We fit the UV continuum in the rest-frame wavelength range of $1200-2100$ \AA\ as well as UV lines of Ly$\alpha$, N \textsc{v}, Si \textsc{iv}, and C \textsc{iv} with stellar and AGN models, assuming that the UV continuum is dominated by stellar and AGN radiation, respectively.

The stellar models consist of the stellar continuum, nebular continuum, nebular emission lines of Ly$\alpha$ and C \textsc{iv}, and ISM/CGM absorption lines of Si \textsc{iv} and C \textsc{iv}, assuming that the continua and emission lines are absorbed by the partially-covering foreground gas in the ISM/CGM. The model flux is expressed as the product of the intrinsic spectrum and a partial-covering absorption factor (e.g., \citealt{Jones2013}) by the following equation:
% \begin{align}
%     f_\lambda&=\left(f_{\lambda}^\mathrm{stellar}+f_{\lambda}^\mathrm{nebular}+\sum_i f_\lambda^{\mathrm{line}_i}\right) \notag\\
%     \times&\left(1-C_f+C_f\exp(-\sum_{j,k}\tau_{j,k}(\lambda))\right),
% \end{align}
\begin{align}
    f_\lambda&=\left(f_{\lambda}^\mathrm{stellar}+f_{\lambda}^\mathrm{nebular}+\sum_i f_\lambda^{\mathrm{line}_i}\right)T_\lambda^\mathrm{ISM+CGM}T_\lambda^\mathrm{IGM}
    %\notag\\\times&\left(1-C_f+C_f\exp(-\sum_{j,k}\tau_{j,k}(\lambda))\right),
\end{align}
where $f_{\lambda}^\mathrm{stellar}$, $f_{\lambda}^\mathrm{nebular}$, $f_{\lambda}^{\mathrm{line}_i}$, $T_\lambda^\mathrm{ISM+CGM}$, and $T_\lambda^\mathrm{IGM}$ represent the \red{stellar continuum, nebular continuum,} nebular emission lines ($i=\mathrm{Ly}\alpha$, C \textsc{iv}), transmission in the ISM/CGM and IGM, respectively. We construct the stellar and nebular continuum models from the stellar population synthesis code BPASS v2.2.1 \citep{Eldridge2017,Stanway2018} and photoionization code \textsc{Cloudy} v23.01 \citep{Ferland1998,Gunasekera2023}, respectively. For the BPASS models, we adopt binary star models, an \citet{Salpeter1955} initial mass function (IMF) with \red{a mass range of $0.1-100\ M_\odot$}, and a bursty star formation history, varying stellar metallicities of $Z_*=10^{-3}-0.040$ and stellar ages of $\log(t/\mathrm{yr})=6.0-11.0$. We assume a \citet{Calzetti2000} extinction law parametrized by a color excess of $E(B-V)$. The model spectra are smoothed to match the observed spectral resolution of $R=1000$. \red{Adopting the BPASS spectra as incident spectra, we calculate the nebular continuum models with \textsc{Cloudy}.} See \citet{Nakane2024b,Nakane2025} for more details of the models. We use a Gaussian function for the emission line as
\begin{align}
f_\lambda^{\mathrm{line}_i}&=\frac{F_i}{\sqrt{2\pi}\sigma}\exp(-\frac{(\lambda-\lambda_{i,0})^2}{2\sigma^2}),
\end{align}
where $F_i$, $\lambda_{i,0}$, and $\sigma$ are the flux, observed wavelength, and velocity dispersion of emission line $i$, respectively.
The transmission in the ISM/CGM is expressed as
\begin{align}
T_\lambda^\mathrm{ISM+CGM}=1-C_f+C_f\exp\left(-\sum_{j,k}\tau_{j,k}(\lambda)\right).
\end{align}
The parameter $C_f$ is the covering fraction, and $\tau_{j,k}(\lambda)$ is the optical depth of the $k$th transition of ion $j$ ($j=\mathrm{Si}$ \textsc{iv}, C \textsc{iv}; $k=1,\ 2$).
The optical depth for the absorbing gas $\tau_{j,k}(\lambda)$ is expressed as
\begin{align}
    \tau_{j,k}(\lambda)&=\tau_{0j,k}\exp(-\frac{(v(\lambda)-v_{0j})^2}{b_D^2}),\\
    \tau_{0j,k}&=\frac{\pi e^2}{m_ec}\frac{\lambda_{0j}f_{j,k}N_{j,k}}{b_D},
\end{align}
where $e$, $m_e$, $c$, $v(\lambda)$, $\tau_{0j,k}$, $v_{0j}$, $b_D$, $\lambda_{0j}$, $f_{j,k}$, and $N_{j,k}$ indicate the elementary charge, mass of electron, speed of light, velocity as a function of wavelength, optical depth and velocity at line center, Doppler width, rest-frame wavelength of lines, oscillator strength, and ionic column density. We use the oscillator strength values shown in \citet{Cashman2017}. For the transmission in the IGM, we use models of \citet{Inoue2014}.

\begin{figure*}[ht!]
    \centering
    \includegraphics[width=0.9\linewidth]{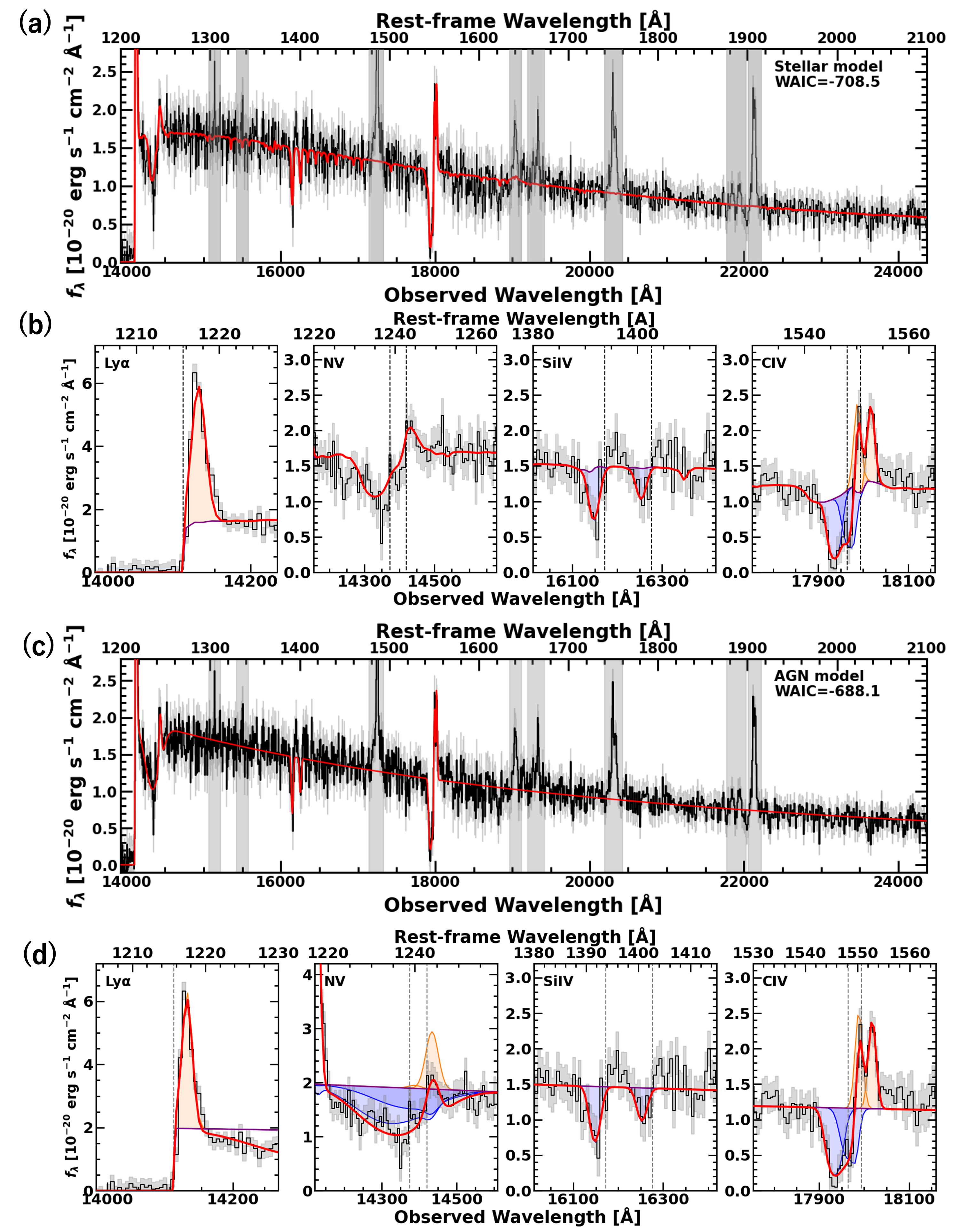}
    \caption{Results of the UV spectral fitting with the stellar (a,b) and AGN (c,d) models. In all panels, the black lines and light gray shaded regions show the MSA spectra and their $1\sigma$ uncertainties, respectively, while the red lines denote the best-fit model spectra. The dark gray shaded regions in panels (a) and (c) represent the emission line wavelength ranges, masked in the fitting.  The panels (b) and (d) present the zoom-in on the Ly$\alpha$, N \textsc{v}, Si \textsc{iv}, and C \textsc{iv} lines. The purple lines, orange shaded regions, and blue shaded regions indicate the best-fit continuum, emission line, and absorption line models, respectively. The black dashed lines show the line wavelengths based on the systemic redshift.}
    \label{fig:UV_fit}
\end{figure*}

The AGN models consist of the AGN continuum, nebular emission lines of Ly$\alpha$, N \textsc{v}, and C \textsc{iv}, and ISM/CGM absorption lines of N \textsc{v}, Si \textsc{iv}, and C \textsc{iv}. The model flux is expressed in a similar way as the stellar models with the following equation:
\begin{align}
    f_\lambda&=\left(f_{\lambda}^\mathrm{AGN}+\sum_i f_\lambda^{\mathrm{line}_i}\right)T_\lambda^\mathrm{ISM+CGM}T_\lambda^\mathrm{IGM},
\end{align}
where $f_\lambda^\mathrm{AGN}$ denotes the AGN power-law continuum of $f_\lambda^\mathrm{AGN}=\alpha\lambda^\beta$.

In our UV spectral fitting, we use the MSA spectra, which provide continuous wavelength coverage redward of the Lyman break. We fit both the stellar and AGN models in the rest-frame wavelength ranges of $1200-2100$ \AA, masking emission and absorption features not included in the models. We conduct MCMC fitting with \texttt{emcee} and compare the stellar and AGN models using the WAIC. The stellar model is strongly favored over the AGN model with \red{$\Delta\mathrm{WAIC}=-20.4$.} The best-fit parameters are summarized in Table \ref{tab:spectral_fitting}. In Figure \ref{fig:UV_fit}, we present the fitting results for the stellar and AGN models. The best-fit stellar model successfully reproduces the P-Cygni profiles through the combination of stellar winds, nebular emission, and ISM/CGM absorption. In contrast, the Si \textsc{iv} and C \textsc{iv} profiles can also be reproduced by the best-fit AGN model through nebular emission and ISM/CGM absorption alone. \red{However, reproducing the N \textsc{v} profile without stellar winds requires a very broad absorption component with an FWHM of $\sim4200$ km s$^{-1}$. Such a width is not unusual by itself for broad-absorption-line quasars (BAL QSOs), in which absorption troughs commonly extend over several thousand km s$^{-1}$ (e.g., \citealt{Weymann1991,Gibson2009}). In our fit, however, a comparably broad absorption component is not required for both C \textsc{iv} and Si \textsc{iv}. In particular, QSO outflows with N \textsc{v} absorption substantially stronger than the corresponding C \textsc{iv} are observed to be very rare \citep{Hamann2018}. The requirement for such an isolated broad N \textsc{v} absorption component therefore makes a simple BAL-like AGN interpretation less natural.}

\begin{table*}
    \centering
    \caption{\red{Best-fit parameters for stellar and AGN models.}}
    \begin{tabular}{lllcc}
    \hline
    \hline
    Component & Parameter & & Stellar model & AGN model \\
    \hline
    Continuum & $\log{(t_\mathrm{age})}$ & [yr] & $6.46_{-0.06}^{+0.03}$ & - \\
     & $\log(Z_*)$ & [$Z_\odot$] & $-1.43_{-0.16}^{+0.13}$ & - \\
     & $E(B-V)$ & & $0.05_{-0.01}^{+0.01}$ & - \\
     & AGN slope $\beta$ & & - & $-2.19_{-0.04}^{+0.03}$ \\
     Nebular emission & $F_\mathrm{Ly\alpha}$ & [$10^{-19}$ erg s$^{-1}$ cm$^{-2}$] & $10.7_{-0.5}^{+0.4}$ & $9.3_{-0.6}^{+0.4}$ \\
     & $\Delta v_\mathrm{Ly\alpha}$ & [km s$^{-1}$] & $453_{-12}^{+9}$ & $450_{-14}^{+7}$  \\
     & FWHM$_\mathrm{Ly\alpha}$ & [km s$^{-1}$] & $478_{-38}^{+15}$ & $409_{-30}^{+22}$ \\
     & $F_\mathrm{NV1238}$ & [$10^{-19}$ erg s$^{-1}$ cm$^{-2}$] & - & $<3.0^\mathrm{a}$ \\
     & $F_\mathrm{NV1243}$ & [$10^{-19}$ erg s$^{-1}$ cm$^{-2}$] & - & $4.5_{-1.6}^{+0.6}$ \\
     & $\Delta v_\mathrm{NV}$ & [km s$^{-1}$] & -  & $275_{-83}^{+83}$ \\
     & FWHM$_\mathrm{NV}$ & [km s$^{-1}$] & - & $822_{-219}^{+219}$ \\
     & $F_\mathrm{CIV1548}$ & [$10^{-19}$ erg s$^{-1}$ cm$^{-2}$] & $2.6_{-0.8}^{+0.8}$  & $2.8_{-0.7}^{+0.8}$ \\
     & $F_\mathrm{CIV1550}$ & [$10^{-19}$ erg s$^{-1}$ cm$^{-2}$] & $2.2_{-0.3}^{+0.3}$  & $2.7_{-0.4}^{+0.1}$ \\
     & $\Delta v_\mathrm{CIV}$ & [km s$^{-1}$] & $394_{-19}^{+19}$  & $402_{-25}^{+16}$ \\
     & FWHM$_\mathrm{CIV}$ & [km s$^{-1}$] & $302_{-39}^{+39}$  & $310_{-41}^{+31}$ \\
     ISM/CGM Absorption & $C_f$ & & $0.88_{-0.15}^{+0.07}$  & $0.92_{-0.14}^{+0.05}$ \\
      & $\log{(N_\mathrm{NV})}$ & [cm$^{-2}$] & -  & $15.6_{-0.1}^{+0.1}$ \\
      & $\Delta v_\mathrm{NV}$ & [km s$^{-1}$] & -  & $-1003_{-232}^{+116}$ \\
      & $b_\mathrm{NV}$ & [km s$^{-1}$] & -  & $2540_{-291}^{+194}$ \\
      & $\log{(N_\mathrm{SiIV})}$ & [cm$^{-2}$] & $14.3_{-0.2}^{+0.1}$  & $14.4_{-0.2}^{+0.1}$ \\
      & $\Delta v_\mathrm{SiIV}$ & [km s$^{-1}$] & $-438_{-61}^{+40}$  & $-445_{-80}^{+30}$ \\
      & $b_\mathrm{SiIV}$ & [km s$^{-1}$] & $264_{-114}^{+68}$ & $271_{-92}^{+50}$ \\
      & $\log{(N_\mathrm{CIV})}$ & [cm$^{-2}$] & $15.2_{-0.2}^{+0.2}$  & $15.2_{-0.1}^{+0.2}$ \\
      & $\Delta v_\mathrm{CIV}$ & [km s$^{-1}$] & $-464_{-41}^{+41}$  & $-501_{-55}^{+37}$ \\
      & $b_\mathrm{CIV}$ & [km s$^{-1}$] & $243_{-50}^{+66}$  & $322_{-90}^{+45}$ \\
    \hline    
    \end{tabular}
    \par
    \vspace{0.02\hsize}
    \footnotesize{a: $3\sigma$ upper limit for a posterior distribution peaking at zero flux.}
    \label{tab:spectral_fitting}
\end{table*}

\subsection{Is the UV Continuum Powered by \\ Massive Stars or an AGN?}
\label{subsec:source}

The empirical comparison with local massive stars (Section \ref{subsec:massive_stars}) and the quantitative spectral fitting (Section \ref{subsec:fitting}) consistently favor a stellar origin of the UV continuum. To further evaluate the possible contribution of an AGN, we combine these results with previous spectroscopic and photometric observations.

\begin{figure}
    \centering
    \includegraphics[width=0.99\linewidth]{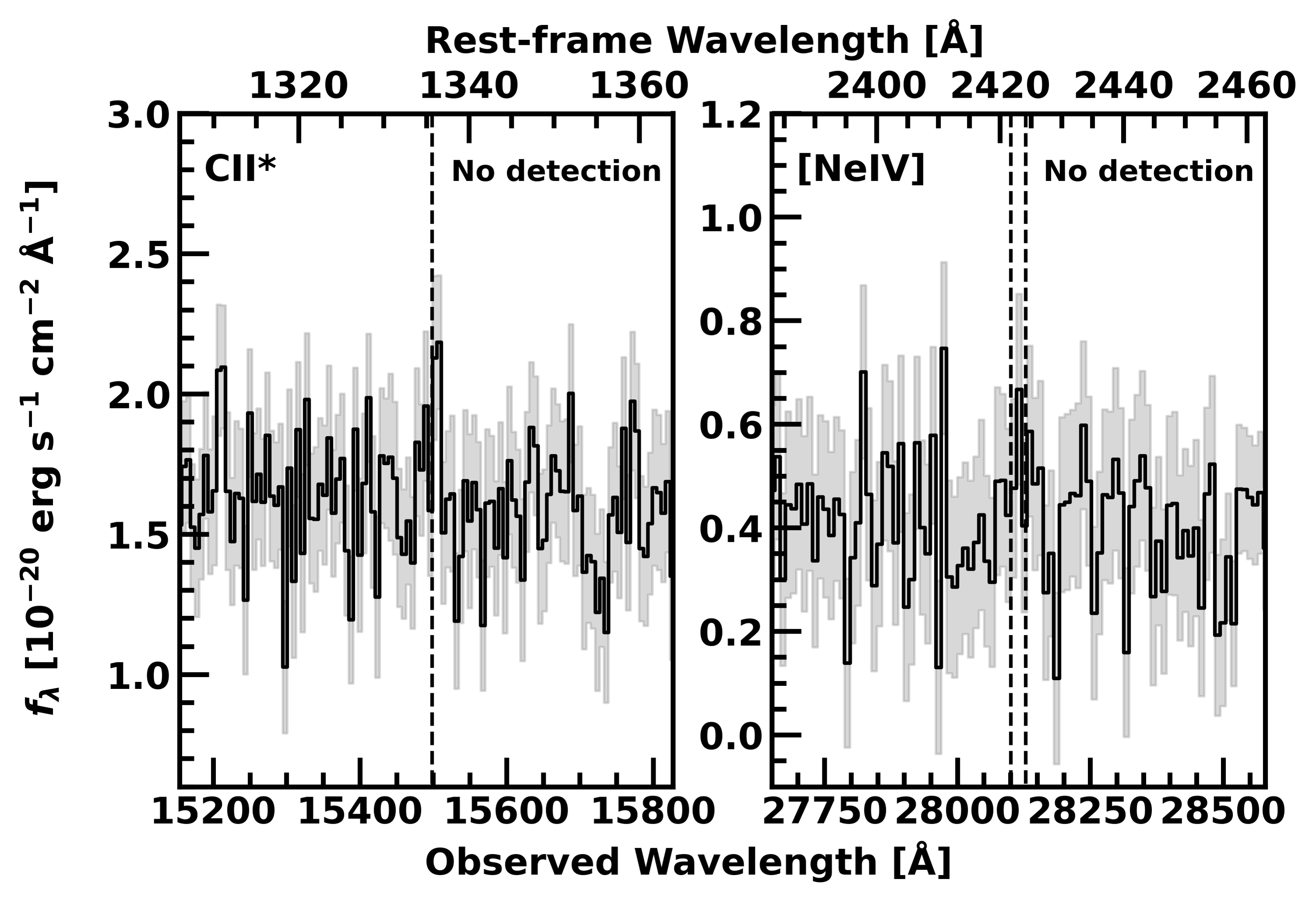}
    \caption{Zoom-in on the MSA spectra around C \textsc{ii}$^*$ and [Ne \textsc{iv}] lines. The black line and gray shaded region represent the observed spectrum and its $1\sigma$ uncertainty, respectively. The black dashed lines indicate the line wavelengths based on the systemic redshift. The MSA spectra do not show any detections of both lines.}
    \label{fig:CII_NeIV}
\end{figure}

While our results strongly suggest that the UV continuum is dominated by massive stars, the origin of the ionizing radiation responsible for the nebular emission remains less clear. We examine the detections of C \textsc{ii}$^*$ $\lambda1335$ and [Ne \textsc{iv}] $\lambda\lambda2422,2424$ doublets, which are used as AGN tracers (e.g., \citealt{Maiolino2024}). As presented in Figure \ref{fig:CII_NeIV}, the C \textsc{ii}$^*$ and [Ne \textsc{iv}] lines are not detected at $\mathrm{S/N}>3$ in the MSA spectra. The [Ne \textsc{iv}] doublet is also covered by the IFU spectrum, but not detected. These non-detections do not support the presence of an AGN, although they do not rule it out. Because the UV emission-line diagnostics alone cannot uniquely distinguish between stellar and AGN ionization, we next consider independent constraints from longer wavelengths.

The rest-frame optical observations obtained with MIRI/MRS provide complementary constraints on the AGN scenario. Analyzing the H$\alpha$ and [O \textsc{iii}] $\lambda5007$ lines, \citet{Alvarez-Marquez2025} detect only narrow components with FWHMs of $231\pm52$ and $189\pm25$ km s$^{-1}$, respectively, although they do not exclude the possible existence of a weak broad H$\alpha$ component associated with the broad line region (BLR) because of the current detection limit. They further investigate the type 1 AGN scenario using low-z AGN relations between the H$\alpha$ and $5100$ \AA\ continuum fluxes \citep{Greene2005} as well as the H$\alpha$ and $2-10$ keV X-ray luminosities \citep{Ho2001,Jin2012}. The predicted $5100$ \AA\ flux is significantly higher than that estimated from the NIRCam imaging \citep{Tacchella2023} and NIRSpec spectrum \citep{Bunker2023}. Likewise, the predicted X-ray luminosity exceeds the $3\sigma$ upper limit derived from the Chandra non-detection \citep{Maiolino2024}. Although they note that the X-ray luminosity of high-z type-1 AGN remains uncertain \citep{Maiolino2024} and that the current upper limit is still compatible with a faint type 2 AGN, these observations do not support a standard type 1 AGN dominating the optical continuum and emission lines.

Additional constraints are provided by the MIRI imaging observations. By combining the NIRCam, NIRSpec, and MIRI spectrophotometry data, \citet{Crespo-Gomez_2026} identify a continuum excess at rest-frame $0.66-0.86\ \mu$m that cannot be explained by mixed stellar populations alone. One possible explanation is hot dust emission associated with the dusty torus around a type 2 AGN. Alternatively, similar continuum excesses have been reported in low-z metal-poor star-forming galaxies \citep{Reines2008,Adamo2010}, where they may arise from hot dust emission associated with dense young clusters or WR stars, or Extended Red Emission (ERE; \citealt{Witt2004}), a photoluminescence mechanism produced by UV-excited small dust grains or complex molecules in the dense circumstellar medium (CSM). The observed spectral energy distribution is not consistent with a standard type-1 AGN, whose optical continuum overwhelms the relatively modest continuum excess observed by MIRI.

To summarize, our empirical comparison with local massive stars and quantitative spectral fitting consistently suggests that the luminous, compact UV continuum of GN-z11 is dominated by massive stars rather than by a classical type 1 AGN. Independent constraints from previous multi-wavelength observations further support this interpretation. However, a contribution from a deeply obscured (type 2) AGN cannot be excluded with the existing observations, particularly in the high-ionization nebular emission and the optical-red continuum excess revealed by the MIRI photometry. Motivated by this preferred stellar interpretation, we next investigate the physical properties of the ionized gas and explore whether the observed broad emission components, stellar-wind signatures, and nitrogen enrichment can be understood within a massive-star scenario.

\section{Gas Properties and Nitrogen Enhancement}
\label{sec:gas_nitrogen}

\subsection{Stellar Winds and Their Connection to Outflows}
\label{subsec:wind_outflow}

\subsubsection{Origin of Broad Components}
\label{subsubsec:broad}

In Section \ref{subsec:UV_feature}, we identify broad-component signatures in N \textsc{iv}] ($\mathrm{FWHM}=1640$ km s$^{-1}$) and He \textsc{ii} ($\mathrm{FWHM}=750$ km s$^{-1}$), which may be associated with stellar winds from massive stars, consistent with the presence of P-Cygni profiles. These line widths indicate gas motions on characteristic scales of several hundred to $\sim1600$ km s$^{-1}$, although they should not be directly equated with the terminal wind velocity. Comparable slow-to-moderate wind \red{velocities} have been observed in local intermediate- to late-type O stars and late WN stars (e.g., \citealt{Prinja1990,Hamann2006}). Such winds may be associated with lower effective temperatures and extended stellar radii, which reduce the stellar escape velocity (e.g., \citealt{Castor1975}) as well as with low metallicity, which weakens line-driven wind acceleration (e.g., \citealt{Leitherer1992,Hawcroft2024}). 

\begin{figure}
    \centering
    \includegraphics[width=0.9\linewidth]{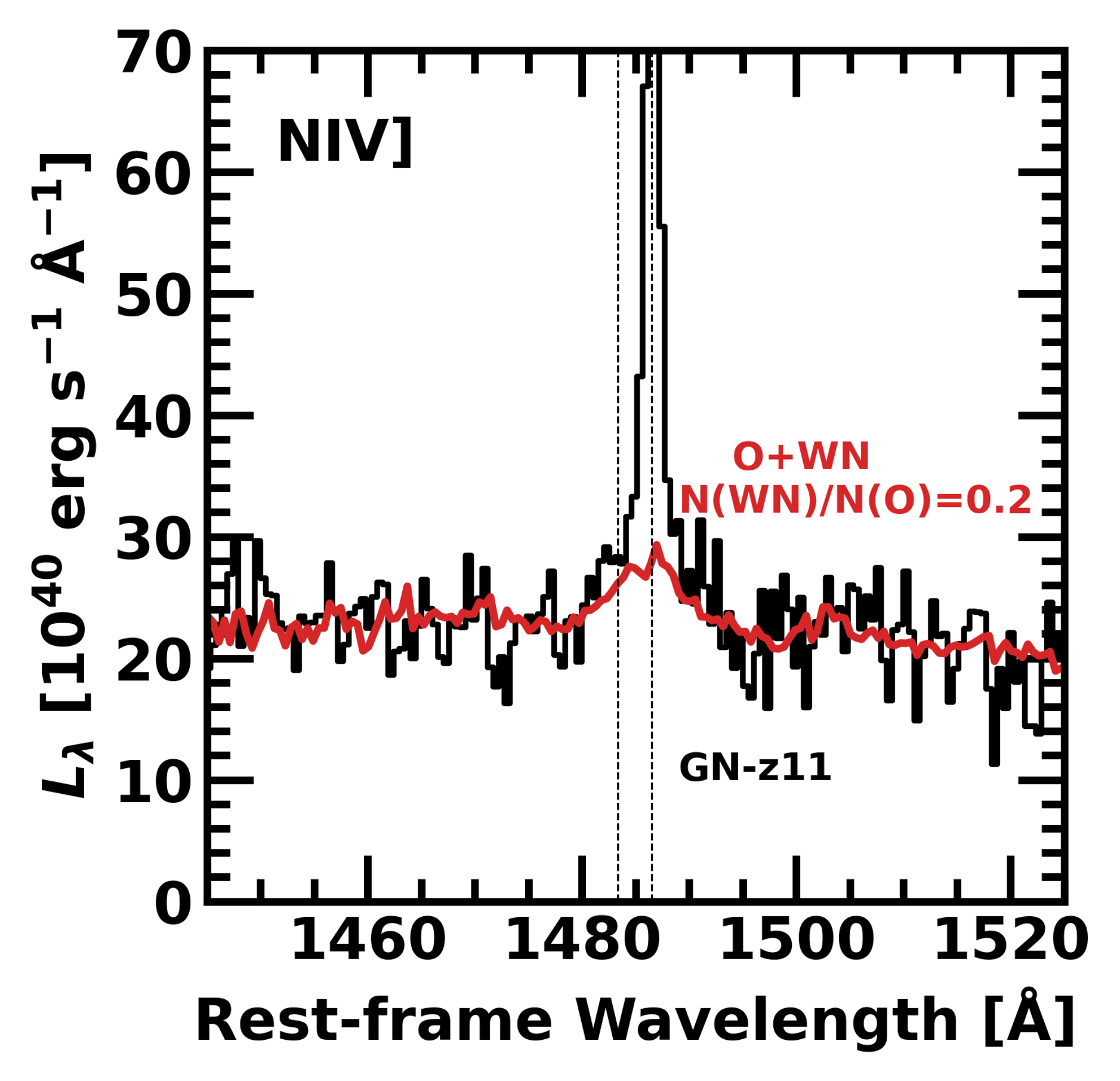}
    \caption{Comparison of N \textsc{iv}] profiles between GN-z11 (black) and O+WN composite (red) spectra. The O+WN composite spectrum is constructed by the linear combination of the observed local O-type and WN stars with the number ratio of N(WN)/N(O)$=0.2$.}
    \label{fig:O+WN}
\end{figure}

To evaluate whether the broad N \textsc{iv}] emission is associated with stellar winds, we compare the N \textsc{iv}] line profile of GN-z11 with that of an empirical O+WN composite spectrum in Figure \ref{fig:O+WN}. The composite spectrum is constructed from the observed ULLYSES spectra of an O 6.5V star (Section \ref{subsec:massive_stars}) and a hydrogen-rich WN6 (WN6h) star, which shows prominent N \textsc{iv}] broad emission. We linearly combine their luminosities assuming a number ratio of $N\mathrm{(WN)}/N\mathrm{(O)}=0.2$. Although this ratio is relatively high, it has been observed in local WR clusters \citep{Hadfield2006}. The composite spectrum reproduces a broad N \textsc{iv}] profile similar to that of GN-z11, suggesting that the broad N \textsc{iv}] emission in GN-z11 is consistent with an origin in the winds of evolved massive stars within the physically possible O/WN number ratios. However, WN stars exhibiting such strong broad N \textsc{iv}] $\lambda1486$ emission are generally accompanied by prominent N \textsc{v}, He \textsc{ii}, and N \textsc{iv} $\lambda1718$ wind features that are much weaker in GN-z11. This suggests that the stellar wind responsible for the broad \red{N \textsc{iv}]} emission may not be represented by a population of classical WN stars. 

\begin{figure}
    \centering
    \includegraphics[width=0.9\linewidth]{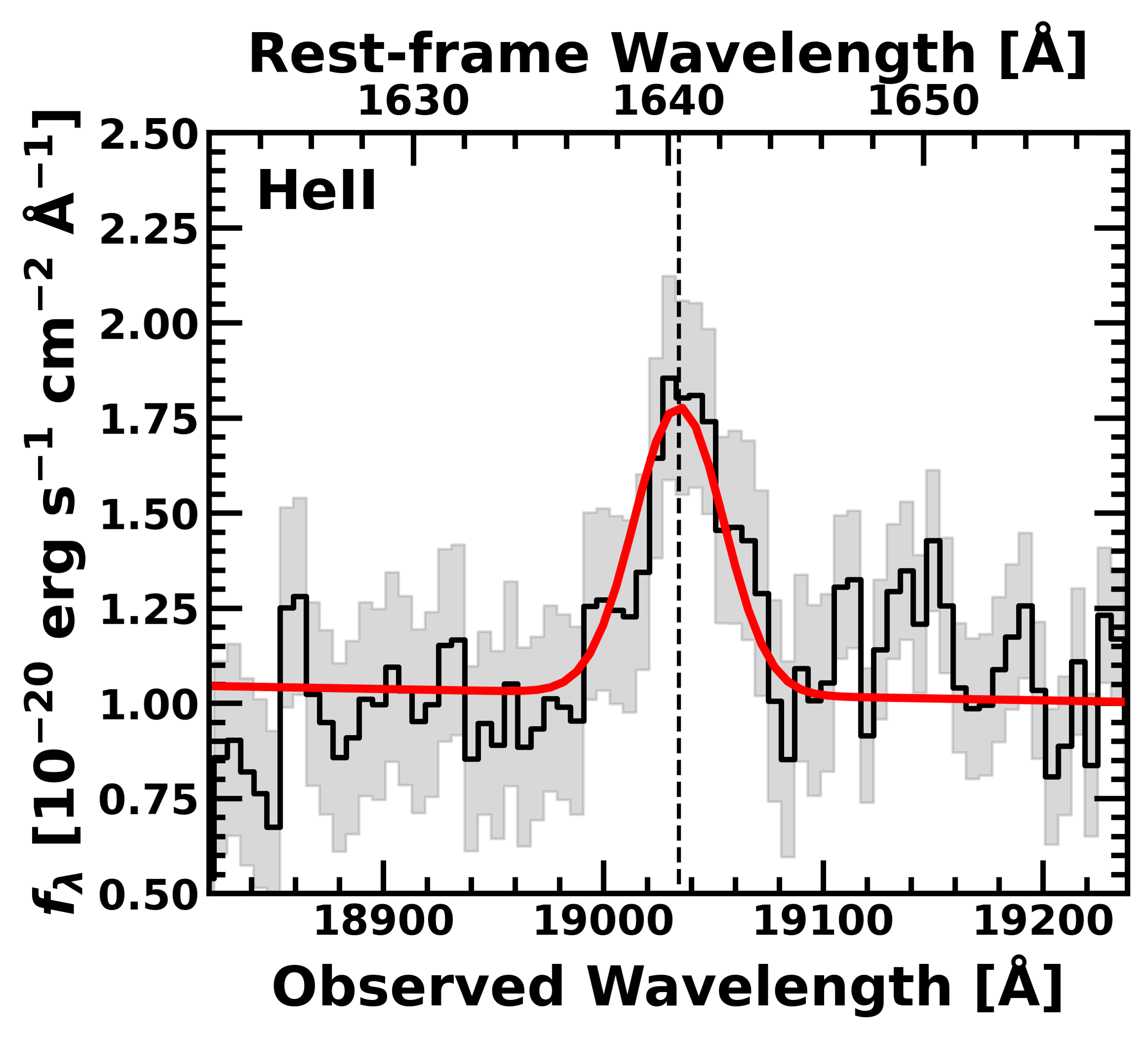}
    \caption{Fitting results for a tentative He \textsc{ii} broad component. The black solid line and shaded region indicate the MSA spectrum of GN-z11 and its $1\sigma$ uncertainty, respectively. The red line shows the best-fit Gaussian function. The estimated line width is FWHM$=752_{-144}^{+144}$ km s$^{-1}$. The black dashed line denotes the line wavelength based on the systemic redshift.}
    \label{fig:HeII}
\end{figure}

The tentative broad He \textsc{ii} of GN-z11 ($\mathrm{FWHM}=750$ km s$^{-1}$; Figure \ref{fig:HeII}) is comparable to RXCJ2248-ID3 with confirmed WR features ($\mathrm{FWHM}=530$ km s$^{-1}$; \citealt{Berg2026}), while significantly broader He \textsc{ii} wind lines have been detected in low-redshift WR galaxies of the Sunburst Arc at $z=2.37$ ($\mathrm{FWHM}=1370$ km s$^{-1}$; \citealt{Rivera-Thorsen2024}) and MARTA-4327 at $z=2.22$ ($\mathrm{FWHM}=1460$ km s$^{-1}$; \citealt{Curti2026}). \citet{Berg2026} claim that the stronger winds in the Sunburst Arc and MARTA-4327 are consistent with their higher metallicities ($12+\log{\mathrm{O/H}}\sim8.5$ and $\sim8.15$, respectively) than RXCJ2248-ID3 ($12+\log{\mathrm{O/H}=7.75}$). This trend may also apply to GN-z11 ($12+\log{\mathrm{O/H}=7.91}$; \citealt{Alvarez-Marquez2025}), suggesting that its broad He \textsc{ii} emission may originate from weak stellar winds of metal-poor WN stars. 

Broad He \textsc{ii} emission has also been reported in the two $z=8.7$ galaxies of CEERS-1019 ($\mathrm{FWHM}=1373$ km s$^{-1}$) and CEERS-1025 ($\mathrm{FWHM}=1154$ km s$^{-1}$) by \citet{Marques-Chaves2026}\red{, using deep UV spectra obtained by the SPURS program}. These galaxies also exhibit substantially stronger N \textsc{v} P-Cygni profiles than GN-z11, and \citet{Marques-Chaves2026} argue that their UV spectra require a contribution from very massive stars (VMSs). In contrast, both the broad He \textsc{ii} emission and the N \textsc{v} P-Cygni profile are less extreme in GN-z11, whose UV spectrum is well reproduced by the BPASS stellar population models (see Figure \ref{fig:UV_fit}) without requiring a dominant VMS contribution.

The preferred stellar age of $\sim3$ Myr (see Table \ref{tab:spectral_fitting}) inferred from the BPASS stellar model fitting naturally coincides with the evolutionary stage at which the first WR stars are expected to appear \red{(e.g., \citealt{Meynet1995})}. This provides a self-consistent picture in which a relatively small population of evolved massive stars contributes to the broad N \textsc{iv}] emission while the UV continuum remains dominated by young massive stars.

\subsubsection{Outflow Signatures}
\label{subsubsec:outflow}

In addition to the broad N \textsc{iv}] and He \textsc{ii} emission, likely originating from dense stellar winds, the best-fit stellar model (Section \ref{subsec:fitting}) provides further insight into the kinematics of the UV spectral features through the decomposition of the stellar-wind, nebular-emission, and ISM/CGM absorption components (see Figure \ref{fig:UV_fit}). 
\red{We identify the red-shifted Ly$\alpha$ and C \textsc{iv} emission together with blue-shifted Si \textsc{iv} and C \textsc{iv} absorption. During the epoch of reionization, Ly$\alpha$ is generally observed redwards of the systemic velocity due to resonant scattering in the surrounding gas and absorption by the intergalactic medium (IGM; e.g., \citealt{Ouchi2020}). We obtain a Ly$\alpha$ velocity offset of $\Delta v=453_{-12}^{+9}$ km s$^{-1}$, supporting the previously reported large velocity offset for GN-z11 \citep{Bunker2023} and remaining relatively large compared with those typically measured for star-forming galaxies at $z\gtrsim7$ (e.g., \citealt{Nakane2024a,Saxena2024,Tang2024,Kageura2025}). Interestingly, the C \textsc{iv} doublet emission is also red-shifted by $\Delta v=394_{-19}^{+19}$ km s$^{-1}$. Red-shifted C \textsc{iv} emission has been reported in a limited number of star-forming galaxies (e.g., \citealt{Vanzella2016,Topping2024}), and resonant scattering has been shown to significantly modify the emergent C \textsc{iv} profiles in local metal-poor galaxies \citep{Senchyna2022}. Thus, the velocity offsets of the Ly$\alpha$ and C \textsc{iv} emission should not be interpreted as direct measurements of the outflow velocity.}

\red{For the absorption components, we infer velocity offsets of $\Delta v=-438_{-61}^{+40}$ km s$^{-1}$ for Si \textsc{iv} and $\Delta v=-464_{-41}^{+41}$ km s$^{-1}$ for C \textsc{iv}, with a shared covering fraction of $C_f=0.88_{-0.15}^{+0.07}$.}
Blue-shifted UV absorption lines have been widely interpreted as signatures of outflowing gas in both local and high-redshift star-forming galaxies (e.g., \citealt{Shapley2003,Steidel2010,Jones2013,VasanGC2026}).
\red{The relatively large covering fraction suggests that the highly ionized outflowing gas covers a substantial fraction of the compact UV-emitting region along the line of sight, although it does not uniquely constrain the three-dimensional geometry of the outflow.
To characterize the high-velocity extent of the absorbing gas, we also estimate the characteristic maximum outflow velocity as $v_\mathrm{out}=|\Delta v|+2\sigma$, where $\Delta v$ and $\sigma$ are velocity offset and dispersion, respectively. This definition has been commonly adopted prescription in studies of galactic outflows (e.g., \citealt{Genzel2011,Wood2015}). This gives $v_\mathrm{out}=775_{-129}^{+172}$ km s$^{-1}$ for Si \textsc{iv} and $805_{-103}^{+77}$ km s$^{-1}$ for C \textsc{iv}. These values lie toward the high-velocity end of outflows observed in star-forming galaxies across a wide range of redshifts, although direct comparisons are subject to differences in tracers and velocity definitions (e.g., \citealt{Heckman2000,Sugahara2017,Sugahara2019,Avery2022,Reichardt_Chu2025,Xu2025}).
}

\red{The Si \textsc{iv} and C \textsc{iv} absorption have line widths of $\sim400-500$ km s$^{-1}$, narrower than the broad He \textsc{ii} emission and especially the broad N \textsc{iv}] component. The large widths of the broad emission components are also qualitatively consistent with the high velocities characteristic of stellar winds in} intermediate- to late-type O stars and late WN stars (e.g., \citealt{Hawcroft2024}). This difference may indicate that the broad emission traces stellar winds in the vicinity of massive stars, whereas the blue-shifted absorption traces outflowing gas on larger scales. One possibility is that momentum and energy injected by stellar winds drive the surrounding ISM outward, producing slower large-scale outflows. Alternatively, radiation pressure from the intense UV radiation field in compact starbursts has also been proposed as an efficient driving mechanism for high-redshift galactic outflows (e.g., \citealt{Ferrara2023,Ferrara2024}). 
\red{In this scenario, galaxies with specific star formation rates (sSFR) exceeding a critical value of $\mathrm{sSFR}\gtrsim25$ Gyr$^{-1}$ can become super-Eddington and drive radiation-pressure-driven outflows \citep{Ferrara2024}. GN-z11 has an estimated sSFR of $\sim25$ Gyr$^{-1}$ \citep{Alvarez-Marquez2025}, placing it near this threshold. The high outflow velocities inferred from the Si \textsc{iv} and C \textsc{iv} absorption are therefore consistent with this scenario, although the relative contributions of stellar winds and radiation pressure cannot be distinguished from the present observations. We note that the decomposition of the stellar-wind, nebular-emission, and ISM/CGM absorption components depends on the adopted stellar continuum models, and the inferred outflow velocities and covering fraction should therefore be interpreted with caution. Nevertheless, the clearly separated narrow C \textsc{iv} emission and absorption doublets are difficult to attribute solely to broad stellar-wind features and likely trace nebular emission and ISM/CGM absorption, respectively.}

\subsection{Electron Density}
\label{subsec:ne}

\subsubsection{Measurement}
\label{subsubsec:ne_measurement}
The UV emission lines of the N \textsc{iv}] $\lambda\lambda1483,1486$ doublet, N \textsc{iii}] $\lambda\lambda1747,1749,1750,1752,1754$ quintet, and C \textsc{iii}] $\lambda\lambda1907,1909$ doublet are widely used as \red{diagnostics of electron density $n_\mathrm{e}$} (e.g., \citealt{Mingozzi2022,Senchyna2024,Topping2024,Maiolino2024,Topping2025a,Topping2025b}), because the flux ratios of N \textsc{iv}] $\lambda1483$/$\lambda1486$, N \textsc{iii}] $\lambda1750$/total, and C \textsc{iii}] $\lambda1909$/$\lambda1907$ depend sensitively on electron density through the different critical densities of the transitions. We derive the electron densities by forward modeling the N \textsc{iv}], N \textsc{iii}], and C \textsc{iii}] multiplets using \texttt{PyNeb}. The continuum is modeled with a power-law function while the narrow emission lines are modeled with Gaussian functions, for which we fix the redshift and line width to that measured from our Gaussian fitting in Section \ref{subsec:UV_feature}. The relative fluxes within each multiplet are calculated with \texttt{PyNeb} as a function of electron density, fixing electron temperature to be that derived from [O \textsc{iii}] $\lambda4363$/$\lambda5007$ ($T_e=\SI{14000}{K}$; \citealt{Alvarez-Marquez2025}). 
In Section \ref{subsec:UV_feature}, we identify a broad N \textsc{iv}] component, which may affect the decomposition of the narrow doublet and the inferred electron density. To evaluate this systematic uncertainty, we consider two fitting approaches for the \red{N \textsc{iv}]} emission: (1) fitting the narrow doublet together with an additional broad Gaussian component; and (2) fitting the narrow doublet together with a stellar \red{N \textsc{iv}]} template constructed from a local WN star. For model (1), the broad component is represented by a single Gaussian function \red{with free line flux and width, and fixed centroid to that of narrow N \textsc{iv}] $\lambda1486$ component.} For model (2), we construct the stellar template from the local WN star, after subtracting the power-law continuum model. We apply the two methods independently to both the MSA and IFU spectra to evaluate the robustness of the inferred electron densities. For both models, the common free parameters are the electron densities of $\log{(n_\mathrm{e}}$(N\textsc{iv}])), $\log{(n_\mathrm{e}}$(N\textsc{iii}])), and $\log{(n_\mathrm{e}}$(C\textsc{iii}])), line fluxes of $F_{1486}$, $F_\mathrm{NIII]}$, and $F_\mathrm{CIII]}$, and power-law parameters of amplitude $\alpha$ and slope $\beta$. We additionally include free parameters of the broad line flux and width for model (1), while we add a free parameter of \red{the normalization of the WN template} $f_\mathrm{WN}$ for model (2). We conduct MCMC fitting with \texttt{emcee} to obtain posterior distributions, adopting flat priors for all the free parameters. The best-fit parameters and their $1\sigma$ uncertainties are determined in the same way as in Section \ref{subsec:UV_feature}.

\begin{figure*}[ht!]
    \centering
    \includegraphics[width=0.99\linewidth]{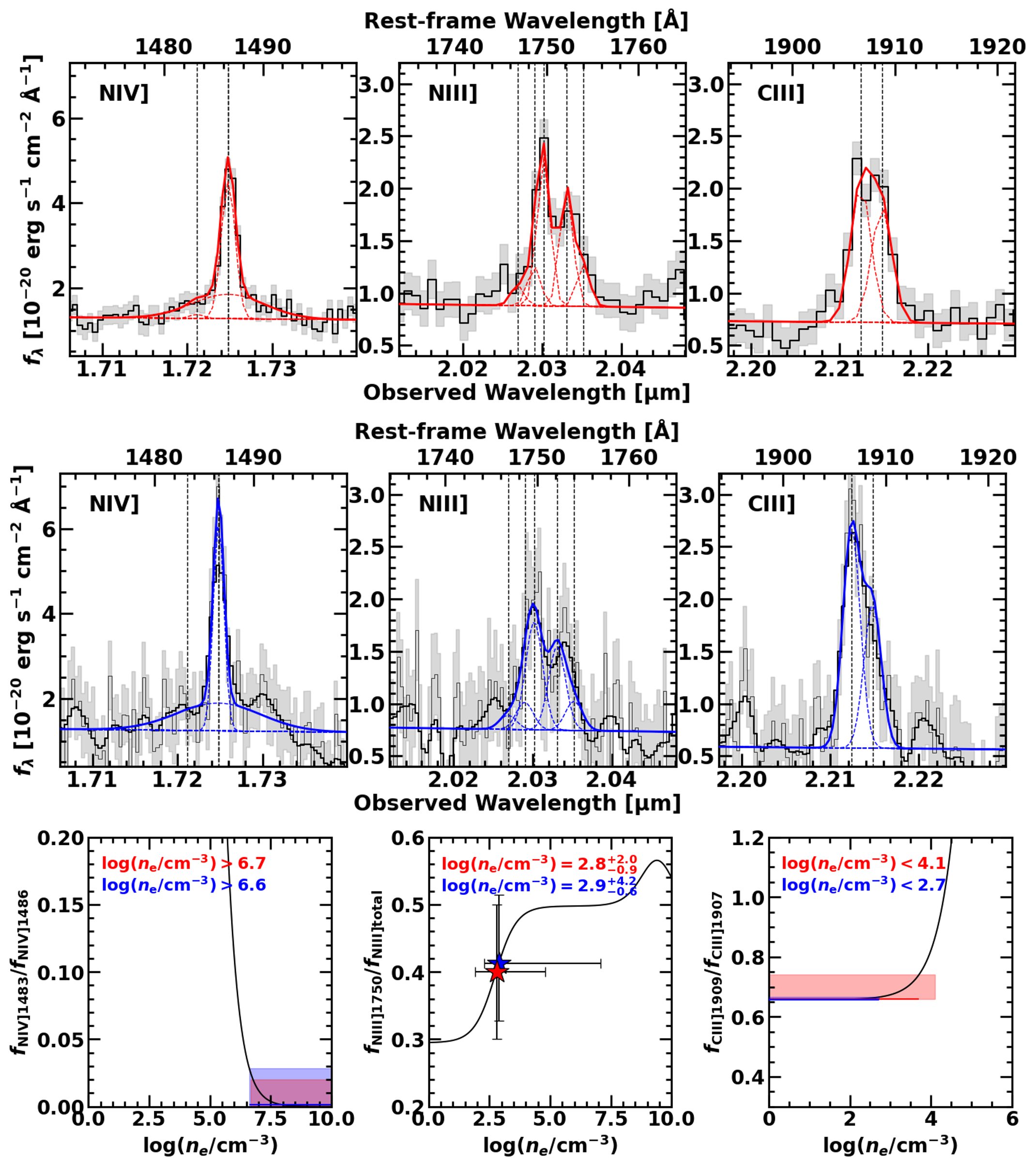}
    \caption{Top: fitting results for the MSA spectrum. The black line and gray shaded region show the observed spectrum and its $1\sigma$ uncertainty, respectively. The red solid and dashed lines represent the best-fit models for all and individual lines, respectively. Middle: same as the top panel, but for the IFU spectrum. Bottom: electron density diagnostics and our measurements. The black lines indicate the relations between the electron density and line ratio calculated with \texttt{PyNeb}, assuming $T_e=14000$ K \citep{Alvarez-Marquez2025}. The red (blue) star symbol and solid lines represent the best-fit measurements while the red (blue) shaded regions indicate the $1\sigma$ uncertainties for the MSA (IFU) spectrum.}
    \label{fig:ne_measurement}
\end{figure*}

\begin{figure*}
    \centering
    \includegraphics[width=0.99\linewidth]{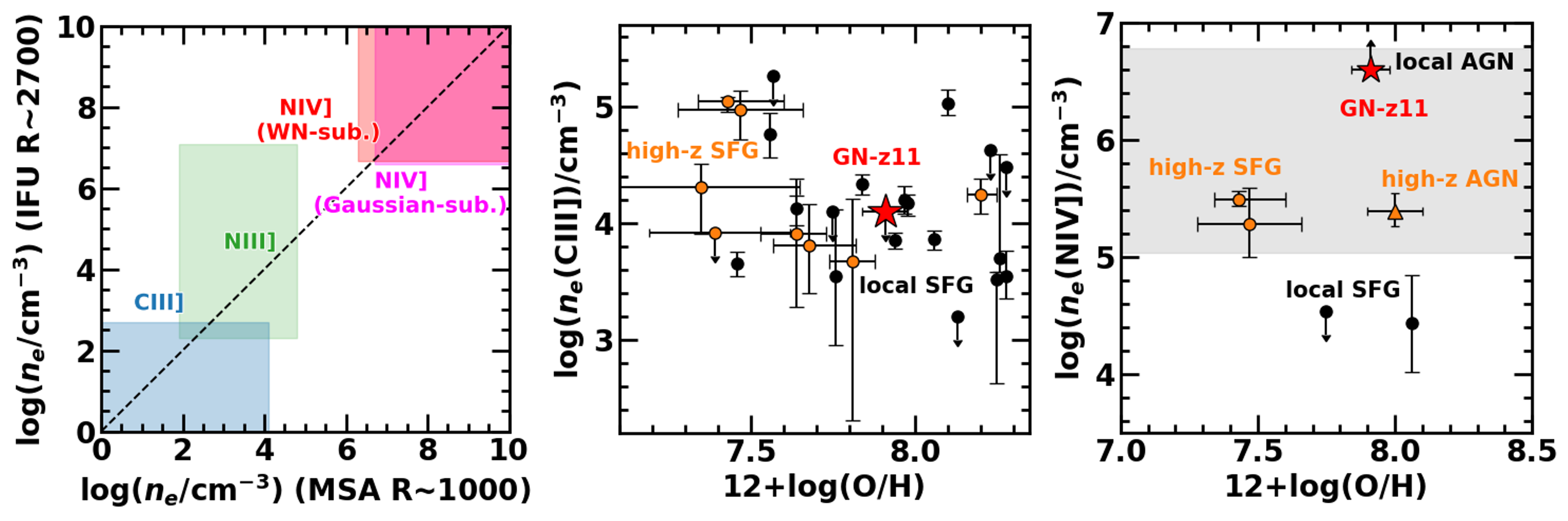}
    \caption{Left: comparison of C \textsc{iii}], N \textsc{iii}], and N \textsc{iv}] electron density measurements between the MSA and IFU spectra. The blue, green, magenta, and red shaded regions show our measurements from C \textsc{iii}], N \textsc{iii}], N \textsc{iv}] (narrow component subtracted by a broad Gaussian component), and N \textsc{iv}] (narrow component subtracted by a N \textsc{iv}] line of the observed WN). The black dashed line indicates that the measurements between MSA and IFU spectra are consistent. Middle: electron densities estimated from C \textsc{iii}] doublet as a function of gas-phase metallicity. The red star symbols show our measurements. The orange and black circles indicate the measurements for high-redshift \citep{Topping2024,Topping2025a,Topping2025b,Sanders2026} and local star-forming galaxies \citep{Mingozzi2022}. Right: same as in the middle panel but for N \textsc{iv}]. The orange triangle and gray shaded region represent the measurements for the high-redshift and local AGN \citep{Ji2024}.}
    \label{fig:ne_results}
\end{figure*}

\red{We further use the oxygen emission lines to constrain the electron densities of the O$^{+}$- and O$^{2+}$-emitting gas. The [O \textsc{ii}] $\lambda3727/\lambda3729$ doublet is sensitive to electron density and has been widely used to measure $n_\mathrm{e}$ in high-redshift galaxies (e.g., \citealt{Isobe2023a,Abdurro'uf2024}). For O$^{2+}$, the combination of O \textsc{iii}] $\lambda1666$, [O \textsc{iii}] $\lambda4363$, and [O \textsc{iii}] $\lambda5007$ enables $T_\mathrm{e}$([O \textsc{iii}]) and $n_\mathrm{e}$([O \textsc{iii}]) to be determined self-consistently without assuming the low-density limit \citep{Hsiao2026}. Following the nebular forward-modeling framework of \citet{Nakane2026}, we simultaneously fit O \textsc{iii}] $\lambda\lambda1661,1666$, [O \textsc{ii}] $\lambda\lambda3727,3729$, H$\delta$, H$\gamma$, and [O \textsc{iii}] $\lambda4363$ in the NIRSpec spectrum with a power-law continuum and Gaussian emission-line profiles, while incorporating the MIRI/MRS H$\alpha$ and [O \textsc{iii}] $\lambda5007$ fluxes of \citet{Alvarez-Marquez2025}. 
To ensure a consistent flux scale between the NIRSpec/MSA and MIRI/MRS measurements, we introduce an additional correction to the co-added MSA spectra, which have already been calibrated to the NIRCam catalog photometry as described in Section \ref{subsec:msa}. Given the residual slit losses associated with the extended UV morphology of GN-z11 reported by \citet{Crespo-Gomez_2026}, we derive a correction factor of $1.55$ by comparing the synthetic F200W photometry from our co-added MSA spectra with the F200W measurement of \citet{Crespo-Gomez_2026}. We apply this factor uniformly to the MSA spectra when combining the NIRSpec and MIRI measurements.
The model has 12 free parameters of $\log n_\mathrm{e}$([O\textsc{ii}]), $T_\mathrm{e}$([O\textsc{ii}]), $\log n_\mathrm{e}$([O\textsc{iii}]), $T_\mathrm{e}$([O\textsc{iii}]), $E(B-V)$, the observed fluxes of H$\alpha$ and [O \textsc{iii}] $\lambda5007$, the intrinsic total [O \textsc{ii}] flux, a common intrinsic velocity width, the fitted redshift, and the amplitude and slope of the power-law continuum. The relative [O \textsc{iii}] line emissivities are calculated with \texttt{PyNeb} as functions of $T_\mathrm{e}$([O\textsc{iii}]) and $n_\mathrm{e}$([O\textsc{iii}]), while the [O \textsc{ii}] doublet ratio is calculated using $T_\mathrm{e}$([O\textsc{ii}]) and $n_\mathrm{e}$([O\textsc{ii}]). The Balmer-line ratios are computed assuming Case B recombination using $T_\mathrm{e}$([O\textsc{iii}]) and $n_\mathrm{e}$([O\textsc{ii}]). We simultaneously model H$\alpha$, H$\gamma$, and H$\delta$ to constrain $E(B-V)$ and apply the corresponding dust attenuation to the nebular lines. We treat $T_\mathrm{e}$([O\textsc{ii}]) as a latent parameter, adopting the $T_\mathrm{e}$([O\textsc{ii}])-$T_\mathrm{e}$([O \textsc{iii}]) relation of \citet{Chakraborty2025} with its model uncertainty as a Gaussian prior, following \citet{Nakane2026}. The observed H$\alpha$ and [O \textsc{iii}] $\lambda5007$ fluxes are likewise treated as latent parameters with Gaussian priors given by the MIRI/MRS measurements of \citet{Alvarez-Marquez2025}. We perform the joint MCMC fitting with \texttt{emcee} to obtain the posterior distributions.}

\begin{figure}
    \centering
    \includegraphics[width=0.99\linewidth]{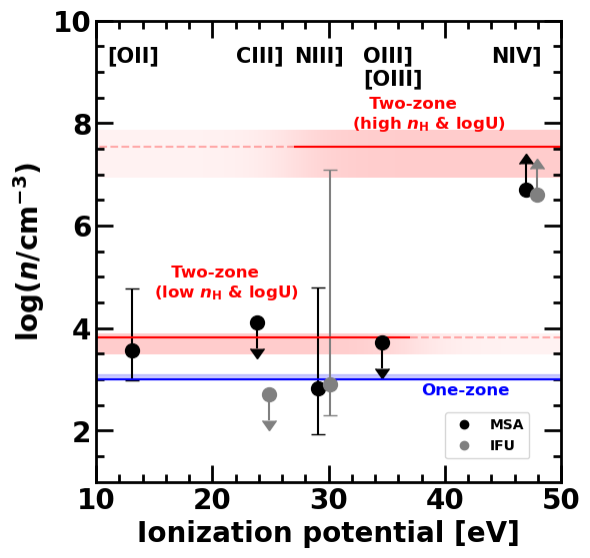}
    \caption{\red{Electron densities traced by multiple ionic species as a function of ionization potential in GN-z11. The black and gray circles represent the constraints obtained from the MSA and IFU spectra, respectively. The blue line and shaded region represent the best-fit hydrogen density and its $1\sigma$ uncertainty obtained from the one-zone Cloudy model. The red lines and shaded regions are the same as the blue ones, but for the two-zone Cloudy model with distinct N/O. Solid and dashed lines indicate the ionization potential ranges over which the respective components are expected to dominate and extend beyond these ranges, respectively.}}
    \label{fig:potential_ne}
\end{figure}

\begin{table*}
\centering
\caption{\red{Electron density constraints of GN-z11.}}
\label{tab:cloudy}
\begin{tabular}{cccc}
\hline\hline
Ionization potential & Line & $\log\ n_\mathrm{e}$ (MSA) & $\log\ n_\mathrm{e}$ (IFU) \\
$[\mathrm{eV}]$ & & [cm$^{-3}$] & [cm$^{-3}$] \\
(1) & (2) & (3) & (4) \\
\hline
$47.5$ & N \textsc{iv}] $\lambda\lambda1483,1486$ & $>6.7$ & $>6.6$ \\
$35.1$ & O \textsc{iii}] $\lambda\lambda1661,1666$, [O \textsc{iii}] $\lambda4363$, [O \textsc{iii}]$\lambda5007$ & $<3.8$ & $-^a$ \\
$29.6$ & N \textsc{iii}] $\lambda\lambda1747-1754$ & $2.8_{-0.9}^{+2.0}$ & $2.9_{-0.6}^{+4.2}$ \\
$24.4$ & C \textsc{iii}] $\lambda\lambda1907,1909$ & $<4.1$ & $<2.7$ \\
$13.6$ & [O \textsc{ii}] $\lambda\lambda3727,3729$ & $3.4_{-0.4}^{+2.1}$ & $-^a$ \\
\hline
\end{tabular}
\par
\vspace{0.02\hsize}
\footnotesize{(1): Ionization potential required to produce the corresponding ions. (2): Emission lines used to constrain the electron density. \\
(3): Electron density constraints from the NIRSpec/MSA spectra and MIRI/MRS [O \textsc{iii}] $\lambda5007$ flux measurements \citep{Alvarez-Marquez2025}. (4): Electron density constraints from the NIRSpec/IFU spectrum.\\
The $1\sigma$ uncertainties and upper/lower limits are presented. $^a$ The required lines are not covered by the IFU spectrum.}
\label{tab:ne}
\end{table*}

\subsubsection{Results and Implications}
\label{subsubsec:ne_results}

In Figure \ref{fig:ne_measurement}, we present the fitting results with model (1), in which the N \textsc{iv}] emission consists of narrow doublet components and a single broad component. The best-fit model spectra are in good agreement with both the MSA and IFU spectra. Since the inferred doublet ratios are located near regions where the diagnostic curves become insensitive to electron density, \red{these} line ratios do not provide unique density measurements but instead constrain $n_\mathrm{e}$ to lower or upper limits. We compare the density measurements among different ions in the left panel of Figure \ref{fig:ne_results}. Regardless of the fitting approaches, the N \textsc{iv}] density reaches a high-density regime of $n_\mathrm{e}\gtrsim10^{6.6}\SI{}{cm^{-3}}$, consistent with previous studies \citep{Senchyna2024,Maiolino2024}. In contrast, the C \textsc{iii}] diagnostic indicates a lower density of $n_\mathrm{e}\lesssim10^{4.1}\SI{}{cm^{-3}}$. The N \textsc{iii}] diagnostic exhibits a wide range of density of $n_\mathrm{e}\sim10^{2-7}$ cm$^{-3}$. As shown in the middle panel of \red{Figure \ref{fig:ne_results}}, our C \textsc{iii}] density measurement is comparable to those measured in local star-forming galaxies (SFGs) from the Cosmic Origins Spectrograph Legacy Spectroscopic Survey (CLASSY; \citealt{Berg2022,Mingozzi2022}) and high-$z$ SFGs at $z\gtrsim5$ \citep{Topping2024,Topping2025a,Topping2025b,Sanders2026}. The N \textsc{iv}] density reaches a higher-density regime than those in local SFGs \citep{Berg2022,Mingozzi2022}, high-$z$ SFGs at $z>6$ \citep{Topping2024,Topping2025a}, and a broad-line AGN at $z=5.55$, GS\_3073 \citep{Ji2024}, as shown in the right panel \red{of Figure \ref{fig:ne_results}}. For reference, we also show the typical $n_\mathrm{e}$(N\textsc{iv}]) range of low-$z$ nitrogen-loud quasars (e.g., \citealt{Jiang2008}), for which we obtain SDSS spectra and derive electron densities by fitting N \textsc{iv}] doublets with double Gaussian functions. \red{Importantly, although the N \textsc {iv}] density of $n_\mathrm{e}\gtrsim10^{6.6}$ cm$^{-3}$ is substantially higher than those measured in some SFGs, it does not require the extreme densities typically associated with AGN BLRs ($n_\mathrm{e}\sim10^{9-11}$ cm$^{-3}$; e.g., \citealt{Rees1989,Peterson2006}). Indeed, similarly high gas densities ($n_\mathrm{e}\sim10^{6-8}$ cm$^{-3}$) can also be found in the environments of massive stars and young star clusters (e.g., \citealt{James2009,Pascale2024,Dai2025}). Likewise, our N \textsc{iii}] diagnostic allows $n_\mathrm{e}\sim10^{2-7}$ cm$^{-3}$ and does not reach $n_\mathrm{e}\gtrsim10^9$ cm$^{-3}$ regime expected for classical BLR gas.} 

\red{The joint oxygen-line modeling yields $T_\mathrm{e}$([O\textsc{iii}])$=16136_{-1478}^{+1055}$ K, which is slightly higher than the previous estimate of $T_\mathrm{e}=14000\pm2100$ K from \citet{Alvarez-Marquez2025}, but consistent within the $1\sigma$ uncertainty. We obtain log$(n_\mathrm{e}$([O\textsc{ii}])/cm$^{-3})=3.4_{-0.4}^{+2.1}$ and log$(n_\mathrm{e}$([O\textsc{iii}])/cm$^{-3})<3.8$. Figure \ref{fig:potential_ne} and Table \ref{tab:ne} summarize the constraints of electron densities from [O \textsc{ii}], C \textsc{iii}], N \textsc{iii}], O \textsc{iii}]/[O \textsc{iii}], and N \textsc{iv}] as a function of the ionization potential required to produce each ion. The combination of UV and optical diagnostics allows us to probe the electron density traced by multiple ionic species within a single galaxy at $z=10.60$. The [O \textsc{ii}], C \textsc{iii}], and [O \textsc{iii}] diagnostics are broadly consistent with relatively low densities, whereas N \textsc{iv}] requires a density higher by several orders of magnitude. The N \textsc{iii}] constraint is substantially broader and also allows higher densities, particularly for the IFU spectrum. The measurements therefore do not indicate a simple monotonic relation between electron density and ionization potential, but instead reveal a strong variation in the densities traced by different ionic species in GN-z11.}

\red{The different electron densities traced by multiple ionic species in GN-z11 suggest that the highly ionized N \textsc{iv}] emission, and possibly a fraction of the N \textsc{iii}] emission, originates from denser gas than the gas traced by [O \textsc{ii}], C \textsc{iii}], and [O \textsc{iii}].} Similar density-stratified structures have been reported in local and high-$z$ galaxies based on the optical/NIR lines (e.g., \citealt{Harikane2025,Usui2025,Takechi2026}) and UV/optical lines (e.g., \citealt{James2009,Mingozzi2022,Ji2024}). A notable stellar example is the local nitrogen-enhanced WR galaxy Mrk996. \citet{James2009} measure an electron density of $n_\mathrm{e}=170\pm40\ \SI{}{cm^{-3}}$ from the narrow [S \textsc{ii}] $\lambda\lambda6717,6731$ ratio, while the broad component reaches $\log{(n_\mathrm{e}/\SI{}{cm^{-3}})}=7.25_{-0.75}^{+1.25}$ based on the O \textsc{iii}] $\lambda1663$/[O \textsc{iii}] $\lambda4363$ and [O \textsc{iii}] $\lambda5007$/$\lambda4363$ ratios. Interestingly, the narrow component shows a normal nitrogen abundance ratio of $\log{(\mathrm{N/O})}=-1.43$ whereas the broad component exhibits a significantly enhanced ratio of $\log{(\mathrm{N/O})}=-0.13$. \red{This indicates that dense and nitrogen-rich gas can be confined to a compact region associated with massive stars without requiring an AGN BLR. In GN-z11, such an interpretation is further supported by the independent signatures of massive stars, including the P-Cygni features and the broad N \textsc{iv}] and He \textsc{ii} emission likely associated with stellar winds. These results therefore suggest that the high-density N \textsc{iv}]-emitting gas may be physically distinct from the more diffuse gas traced by C \textsc{iii}] and the optical oxygen lines. In the following section, we investigate whether such density stratification, together with possible variations in chemical abundances, can account for the observed UV and optical emission-line fluxes using photoionization models.}

\subsection{\red{Photoionization Modeling}}
\red{We model the UV-to-optical emission-line fluxes using the photoionization code \textsc{Cloudy} v23.01 to investigate whether the density stratification inferred in Section \ref{subsec:ne} can reproduce the observed emission lines in the similar way as \citet{Torralba2026}. We adopt the best-fit BPASS spectrum with $\log(t_\mathrm{age}/\mathrm{yr})=6.5$ and $\log(Z_*/Z_\odot)=-1.4$ (see Section \ref{subsec:fitting} and Table \ref{tab:spectral_fitting}) as the incident ionizing spectrum, assuming dust-free gas, constant gas density, plane-parallel geometry, and a covering factor of unity. We calculate grid models by varying the gas-phase metallicity ($\mathrm{[O/H]}=-1.25$ to $-0.25$ in $0.25$ dex steps), hydrogen number density ($\log(n_\mathrm{H}/\mathrm{cm^{-3}})=3.0-8.0$ in $0.5$ dex steps), ionization parameter ($\log U=-2.5$ to $0.0$ in $0.5$ dex steps), nitrogen abundance ($\mathrm{[N/O]}=-1.0$ to $1.5$ in $0.5$ dex steps), and carbon abundance ($\mathrm{[C/O]}=-1.5$ to $0.0$ in $0.5$ dex steps). We fit the NIRSpec/MSA spectra by forward modeling the UV CNO (N \textsc{iv}], O \textsc{iii}], N \textsc{iii}], and C \textsc{iii}] multiplets), optical oxygen ([O \textsc{ii}] doublet and [O \textsc{iii}] $\lambda4363$), and Balmer (H$\delta$ and H$\gamma$) emission lines, while incorporating the MIRI/MRS H$\alpha$ and [O \textsc{iii}] $\lambda5007$ flux measurements of \citet{Alvarez-Marquez2025}. We note that the C \textsc{iv} doublet lines is excluded from our photoionization modeling because their observed fluxes may be affected by stellar-wind emission.  We treat the observed H$\gamma$ flux as a free parameter and calculate the fluxes of the other emission lines from the \textsc{Cloudy}-predicted line ratios relative to H$\gamma$, accounting for dust attenuation with $E(B-V)$. The emission lines are modeled with Gaussian profiles with a common intrinsic velocity width and fitted redshift, convolved with the LSF, together with a power-law continuum. For the broad N \textsc{iv}] component, we fix its flux and FWHM to the best-fit values obtained in Section \ref{subsec:UV_feature}, while tying its centroid to the fitted redshift at the rest wavelength of N \textsc{iv}] $\lambda1486$.}

\begin{figure*}
    \centering
    \includegraphics[width=0.99\linewidth]{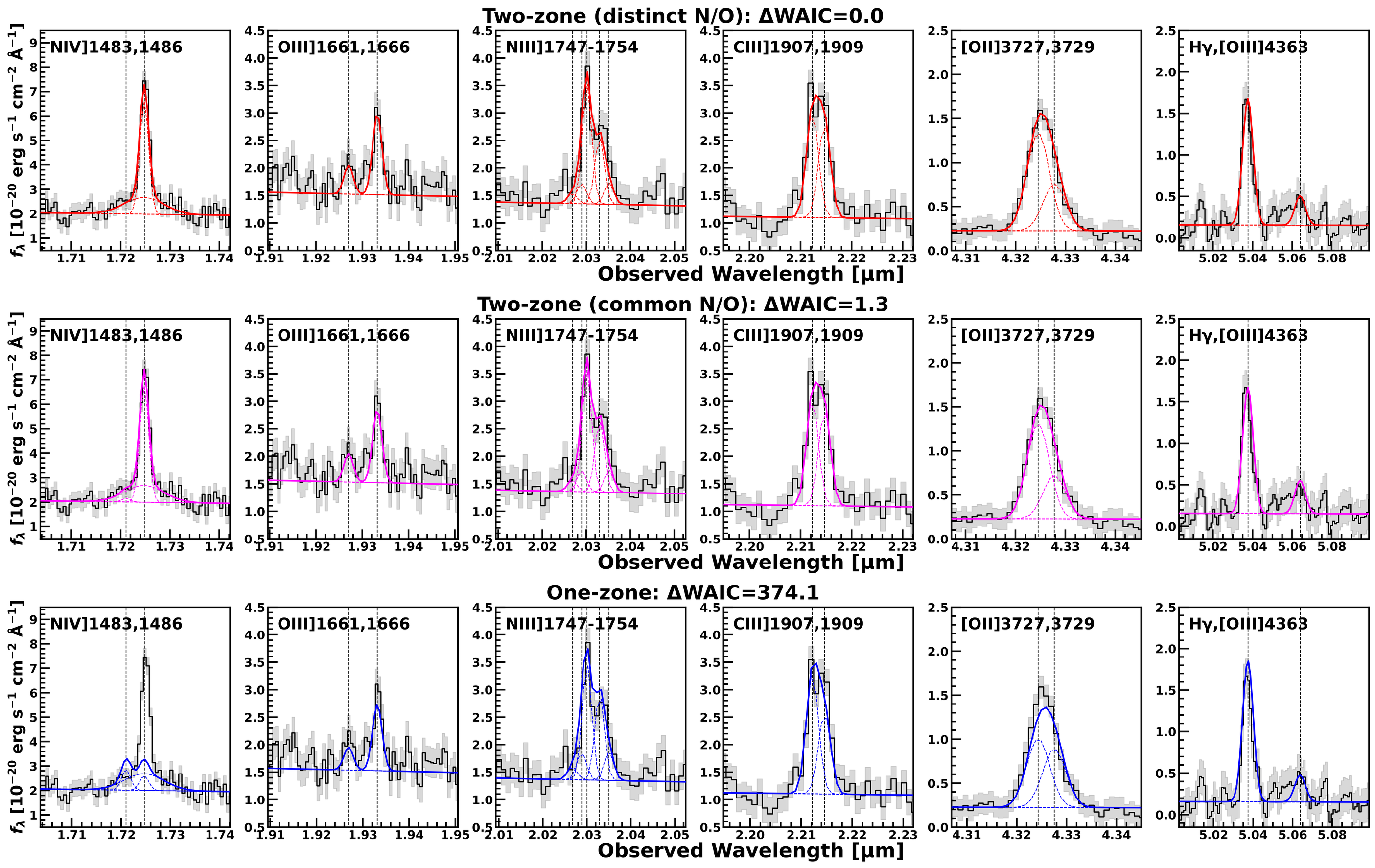}
    \caption{\red{Best-fit \textsc{Cloudy} models for the MSA spectra. The top, middle, and bottom rows show the two-zone model with distinct N/O, the two-zone model with common N/O, and the one-zone model, respectively. The black lines and gray shaded regions represent the observed spectra and their $1\sigma$ uncertainties, respectively. The red, magenta, and blue lines indicate the best-fit model spectra for the three model sets. The corresponding $\Delta\mathrm{WAIC}$ values relative to the model with the lowest WAIC are indicated above each row.}}
    \label{fig:cloudy_spectra}
\end{figure*}

\red{We consider three model configurations: (1) a one-zone model with a single set of physical parameters, (2) a two-zone model with a common N/O abundance ratio, and (3) a two-zone model with distinct N/O abundance ratios. In the two-zone models, the high- and low-density zones have independent values of $n_\mathrm{H}$, $U$, O/H, and C/O, while N/O is either shared or allowed to vary independently between the two zones. The predicted emission-line fluxes from the two zones are combined to construct the total model spectrum as
$(F_\mathrm{line}/F_\mathrm{H\gamma})_\mathrm{total}=(1-w)(F_\mathrm{line}/F_\mathrm{H\gamma})_\mathrm{low\ \textit{n}_H}+w(F_\mathrm{line}/F_\mathrm{H\gamma})_\mathrm{high\ \textit{n}_H}$, where $w$ represents the fraction of the total H$\gamma$ flux contributed by the high-density zone. We adopt flat priors over the model grids and perform MCMC fitting with \texttt{emcee}. The NIRSpec spectrum is fitted using Gaussian likelihoods based on the spectral uncertainties, while the MIRI/MRS H$\alpha$ and [O \textsc{iii}] $\lambda5007$ measurements are incorporated through Gaussian likelihoods using the fluxes and uncertainties reported by \citet{Alvarez-Marquez2025}. We compare the models using the WAIC, where a lower value indicates a preferred model.}

\begin{figure*}
    \centering
    \includegraphics[width=0.8\linewidth]{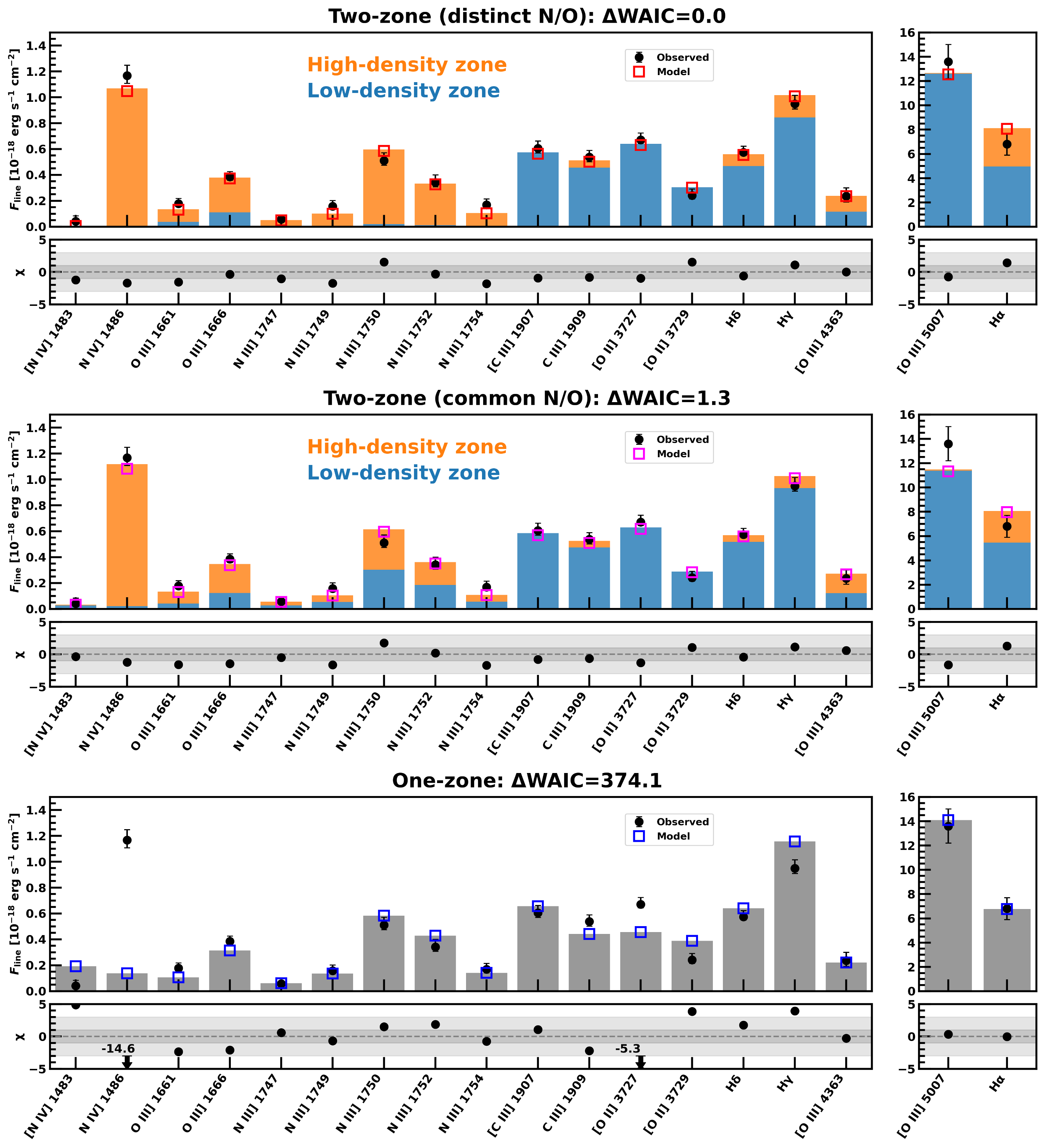}
    \caption{\red{Comparison of the observed and predicted emission-line fluxes for the \textsc{Cloudy} models. The top, middle, and bottom panels show the two-zone model with distinct N/O, the two-zone model with common N/O, and the one-zone model, respectively. The black circles with error bars indicate the observed fluxes and their $1\sigma$ uncertainties, while the open squares represent the best-fit model predictions. For the two-zone models, the orange and blue bars show the contributions from the high- and low-density zones, respectively. The lower panels show the normalized residuals, $\chi=(F_\mathrm{model}-F_\mathrm{obs})/\sigma_\mathrm{obs}$. The dark and light gray shaded regions denote the $1\sigma$ and $3\sigma$ intervals, respectively. The right panels show [O \textsc{iii}] $\lambda5007$ and H$\alpha$ on a separate flux scale.}}
    \label{fig:cloudy_contribution}
\end{figure*}

\red{Figure \ref{fig:cloudy_spectra} presents the best-fit spectra for the three models, while Figure \ref{fig:cloudy_contribution} compares the observed and predicted emission-line fluxes. The best-fit parameters and their uncertainties are summarized in Table \ref{tab:cloudy}. The one-zone model fails to reproduce the observed UV-to-optical line fluxes simultaneously. Particularly, N \textsc{iv}] $\lambda1486$ and [O \textsc{ii}] $\lambda3727$ fluxes are highly underestimated. In contrast, both two-zone models provide substantially improved fits to the observed emission lines. The one-zone model is strongly disfavored relative to the two-zone model with distinct N/O, with $\Delta\mathrm{WAIC}=374.1$. The two-zone models with common and distinct N/O yield nearly identical fitting results, with a small difference of $\Delta\mathrm{WAIC}=1.3$. These results indicate that at least two gas components with different densities and ionization parameters are required to reproduce the observed emission lines, with the high-density zone ($n_\mathrm{H}\sim10^{7.5}\ \mathrm{cm^{-3}}$) exhibiting a higher ionization parameter ($\log U\sim-1.0$) in both two-zone models. Distinct N/O ratios between the components are slightly favored, although the difference is not statistically significant. The best-fit chemical abundances also differ between the two gas components. In the two-zone model with distinct N/O, the low- and high-density zones have oxygen abundances of $12+\log({\rm O/H})=8.20_{-0.12}^{+0.07}$ and $7.94_{-0.18}^{+0.05}$, respectively, with the high-density zone exhibiting a lower metallicity. The corresponding abundance ratios are $\mathrm{[N/O]}=-0.89_{-0.09}^{+1.16}$ and $1.14_{-0.14}^{+0.08}$, and $\mathrm{[C/O]}=-0.02_{-0.10}^{+0.02}$ and $-1.49_{-0.00}^{+0.49}$, indicating higher N/O and lower C/O in the high-density zone. The two-zone model with common N/O shows a similar trend in O/H and C/O, with a shared abundance ratio of $\mathrm{[N/O]}=0.94_{-0.05}^{+0.03}$.}

\begin{table*}
\centering
\caption{\red{Best-fit parameters of the Cloudy photoionization models for GN-z11.}}
\label{tab:cloudy}
\begin{tabular}{lccccc}
\hline\hline
& One-zone
& \multicolumn{2}{c}{Two-zone (common N/O)}
& \multicolumn{2}{c}{Two-zone (distinct N/O)} \\
\cline{3-4}\cline{5-6}
Parameter &
& Low-density zone & High-density zone
& Low-density zone & High-density zone \\
\hline
$\log(n_{\rm H}/{\rm cm}^{-3})$
& $3.01_{-0.00}^{+0.09}$ & $3.61_{-0.21}^{+0.19}$ & $7.58_{-0.56}^{+0.16}$ & $3.80_{-0.33}^{+0.13}$ & $7.46_{-0.48}^{+0.32}$ \\
$\log U$
& $-2.01_{-0.05}^{+0.01}$ & $-2.24_{-0.15}^{+0.06}$ & $-0.99_{-0.24}^{+0.50}$ & $-2.20_{-0.19}^{+0.03}$ & $-1.33_{-0.20}^{+0.64}$ \\
$12+\log({\rm O/H})$
& $7.94_{-0.03}^{+0.02}$ & $8.19_{-0.11}^{+0.03}$ & $7.94_{-0.10}^{+0.07}$ & $8.20_{-0.12}^{+0.07}$ & $7.94_{-0.18}^{+0.05}$ \\
$\mathrm{[N/O]}$
& $0.95_{-0.05}^{+0.03}$ & \multicolumn{2}{c}{$0.94_{-0.05}^{+0.03}$} & $-0.89_{-0.09}^{+1.16}$ & $1.14_{-0.14}^{+0.08}$ \\
$\mathrm{[C/O]}$
& $-0.32_{-0.06}^{+0.01}$ & $-0.02_{-0.08}^{+0.01}$ & $-1.02_{-0.47}^{+0.07}$ & $-0.02_{-0.10}^{+0.02}$ & $-1.49_{-0.00}^{+0.49}$ \\
$E(B-V)$
& $0.00_{-0.00}^{+0.01}$ & \multicolumn{2}{c}{$0.00_{-0.00}^{+0.01}$} & \multicolumn{2}{c}{$0.00_{-0.00}^{+0.01}$} \\
$w^\dagger$
& $-$ & \multicolumn{2}{c}{$0.11_{-0.02}^{+0.02}$} & \multicolumn{2}{c}{$0.11_{-0.02}^{+0.04}$} \\
\hline
WAIC
& $-3167.0$ & \multicolumn{2}{c}{$-3539.8$}
& \multicolumn{2}{c}{$-3541.1$} \\
\hline
\end{tabular}
\par
\vspace{0.01\hsize}
\footnotesize{The uncertainties represent the 68\% HPDI. \\ $^\dagger$ This parameter represents the fraction of the total H$\gamma$ flux contributed by the high-density zone.}
\end{table*}

\red{As shown in Figure \ref{fig:cloudy_contribution}, both two-zone models indicate that N \textsc{iv}] emission is dominated by the high-density zone, whereas C \textsc{iii}] and [O \textsc{ii}] emission predominantly originate from the low-density zone, with little contribution from the high-density zone. In contrast, both components contribute to the O \textsc{iii}] $\lambda\lambda1661,1666$ and [O \textsc{iii}] $\lambda4363$ emission. 
As shown in Figure \ref{fig:potential_ne}, the hydrogen densities inferred from both two-zone models are broadly consistent with the electron density constraints from individual emission lines, which indicate substantially higher densities for N \textsc{iv}] than for C \textsc{iii}] and the optical oxygen lines. 
The contribution to N \textsc{iii}] emission differs between the two-zone models. The high-density zone dominates in the model with distinct N/O, whereas the low-density component contributes substantially in the model with common N/O. This ambiguity is consistent with the broad density range allowed by the N \textsc{iii}] diagnostic in Section \ref{subsubsec:ne_results} (see also Appendix \ref{Appendix:N3}). Thus, although the photoionization models support a multi-component gas structure, the relative nitrogen abundances of the components remain unknown. We discuss the implications for nitrogen enrichment in the following section.}

\subsection{Nitrogen Enhancement}
\label{subsec:nitrogen}

As described in Section \ref{sec:introduction}, the nitrogen abundance ratio of GN-z11 is as high as $\log{\mathrm{(N/O)}}>-0.25$ ($\mathrm{[N/O]}>0.58$) based on the N \textsc{iv}] and N \textsc{iii}] \citep{Cameron2023}, which is much higher than those in local galaxies (e.g., \citealt{Izotov2006,Berg2021}) and Galactic H \textsc{ii} regions \citep{Garcia-Rojas2007} at similar metallicity. Our \red{electron-density measurements and photoionization modeling indicate that the narrow N \textsc{iv}] emission predominantly arises from a high-density, high-ionization gas, distinct from the lower-density, lower-ionization gas traced by C \textsc{iii}] and the optical oxygen lines. These results raise the possibility that the extreme nitrogen abundance inferred from the UV nitrogen lines may not represent the galaxy-wide chemical composition of GN-z11.}

\begin{figure}
    \centering
    \includegraphics[width=0.99\linewidth]{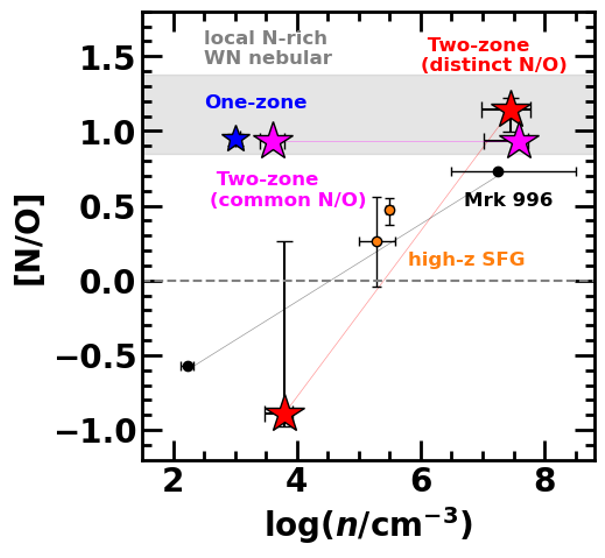}
    \caption{\red{Gas density and nitrogen-to-oxygen abundance ratio. The blue star symbol represents the one-zone \textsc{Cloudy} model, while the magenta and red star symbols connected by lines represent the low- and high-density zones of the two-zone models with common and distinct N/O, respectively. The black circles connected by a gray line indicate the diffuse and dense gas components of the local WR galaxy Mrk 996 \citep{James2009}, while the orange circles represent high-redshift star-forming galaxies \citep{Topping2024,Topping2025a}. The gray shaded region indicates the range of nitrogen abundances in local nitrogen-rich WN nebulae \citep{Fernandez-Martin2012,Vamvatira-Nakou2016,Sevigny2021}. The horizontal dashed line indicates the solar N/O ratio.}}
    \label{fig:ne_NO}
\end{figure}

\red{Figure \ref{fig:ne_NO} compares the gas densities and nitrogen abundances inferred from our photoionization models with those of local and high-redshift nitrogen-enhanced galaxies. In the two-zone model with distinct N/O, the high-density component exhibits enhanced N/O, while the low-density component has substantially lower N/O. This trend qualitatively resembles that observed in the local WR galaxy Mrk 996 \citep{James2009}, where the dense component has a significantly higher N/O ratio than the surrounding diffuse gas. The N/O ratio of the high-density component is also comparable to those found in local nitrogen-rich WN nebulae \citep{Fernandez-Martin2012,Vamvatira-Nakou2016,Sevigny2021}. While the two-zone model with common N/O provides a similarly good fit ($\Delta\mathrm{WAIC}=1.3$), indicating that the evidence for distinct nitrogen abundances between the two zones is not statistically significant, these similarities suggest that localized nitrogen enrichment associated with massive stars remains a plausible explanation for the dense, nitrogen-emitting gas in GN-z11.}

\red{Recent studies of local star-forming galaxies provide further evidence for a connection between nitrogen enrichment and dense star-forming environments. \citet{Arellano-Cordova2025} investigate 45 galaxies from the CLASSY survey \citep{Berg2022} and find that elevated N/O is associated with higher electron densities, compact star formation, and high-velocity outflows. Intriguingly, GN-z11 also exhibits high-velocity ionized-gas outflows, as discussed in Section \ref{subsubsec:outflow}. \citet{James2026} further compare abundances of neutral and ionized gas in 31 CLASSY galaxies and find that N/O in the ionized gas is systematically higher by $\sim0.7$ dex than that in the neutral gas, while substantially smaller offsets between the neutral and ionized gas are found for S/O and Fe/O. They also find that the N/O offset increases rapidly within $3-6$ Myr after the onset of the current starburst, consistent with prompt nitrogen enrichment by WR stars rather than AGB stars. Although these studies use the low-ionization N \textsc{i} and [N \textsc{ii}] lines, the phase-dependent nitrogen abundance and its connection with density and outflows provide empirical support for localized nitrogen enrichment associated with massive stars and stellar feedback.}

\begin{figure*}
    \centering
    \includegraphics[width=0.99\linewidth]{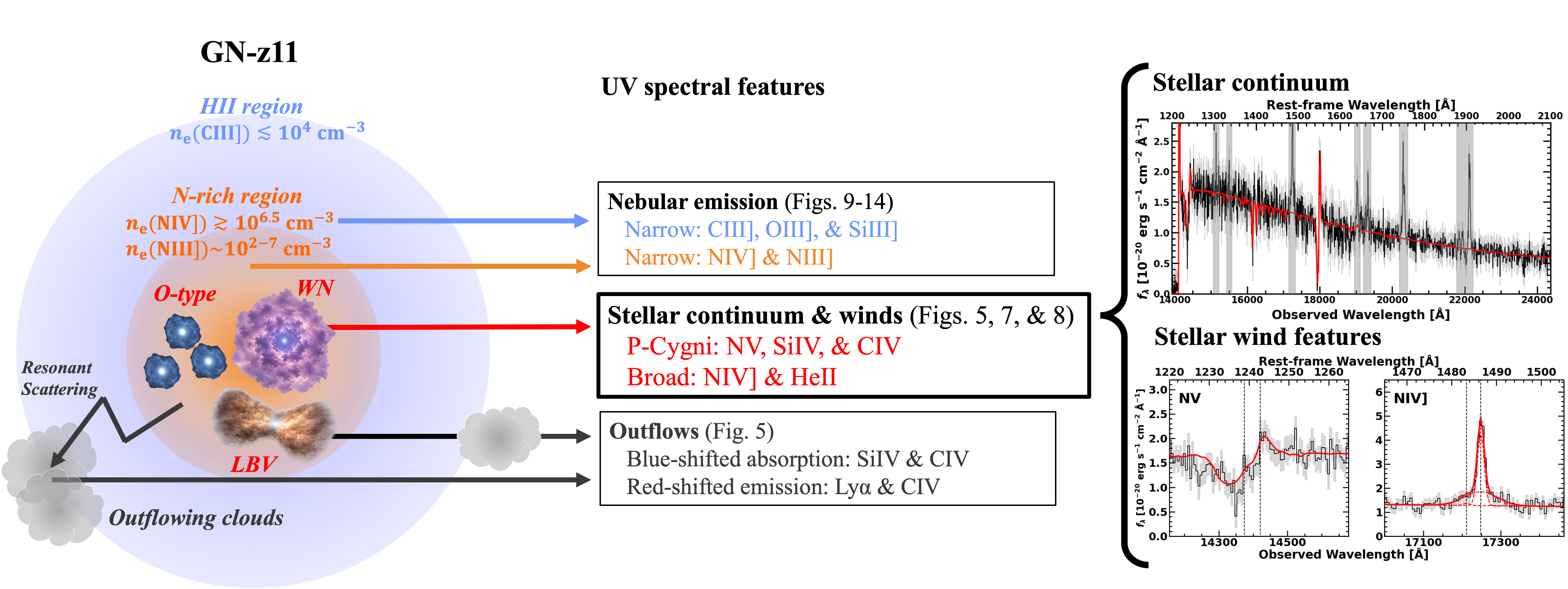}
    \caption{Schematic view of GN-z11 illustrating the proposed origin of the observed UV spectral features. Massive stars, including O-type stars, WN stars, and LBVs, produce a bright UV continuum and stellar winds (red line), which may give rise to the observed P-Cygni profiles (N \textsc{v}, Si \textsc{iv}, and C \textsc{iv}) and broad emission component (N \textsc{iv}] and He \textsc{ii}). The dense gas ($n_\mathrm{e}\gtrsim10^{6.5}$ cm$^{-3}$) surrounding the massive stars is locally enriched in nitrogen by nitrogen-rich stellar winds and photoionized by the nearby massive stars, producing prominent N \textsc{iv}] and possibly N \textsc{iii}] emission (orange line). The other narrow UV emission lines (C \textsc{iii}], O \textsc{iii}], and Si \textsc{iii}]) may arise from a more diffuse H \textsc{ii} region with a moderate density $n_\mathrm{e}\lesssim10^4$ cm$^{-3}$ (blue lines). Outflowing gas clouds may produce the blue-shifted absorption (Si \textsc{iv} and C \textsc{iv}) through foreground absorption while resonant scattering gives rise to the red-shifted emission (Ly$\alpha$ and C \textsc{iv}; black line).}
    \label{fig:structure}
\end{figure*}

\red{Similar evidence has recently emerged at high redshift. \citet{Ji2024} find that GS\_3073 exhibits distinct high-density, high-N/O gas traced by the UV N \textsc{iv}] emission and lower-density, lower-N/O gas traced by optical [N \textsc{ii}] emission. \citet{Umeda2026} report that, in stacked spectra of low-mass galaxies at $z=4.5-10.1$ with $M_*\simeq10^{7.7}\ M_\odot$, N/O inferred from N \textsc{iv}] is $\sim1.4$ dex higher than that inferred from [N \textsc{ii}]. Together with the strong He \textsc{ii} emission, they discuss a possible contribution from WR stars to the hard ionizing radiation and localized nitrogen enrichment. Importantly, their photoionization models show that adopting a hard WR ionizing spectrum alone does not reproduce the strong UV nitrogen emission at normal N/O, and an enhanced nitrogen abundance is still required (see also \citealt{Gunawardhana2025}). More recently, \citet{Torralba2026} identify a similar density-stratified structure in COLA1 at $z=6.6$, with $n_\mathrm{e}\gtrsim10^{6.5}$ cm$^{-3}$ from N \textsc{iv}] and $n_\mathrm{e}\lesssim10^{3.5}$ cm$^{-3}$ from C \textsc{iii}]. Their photoionization analysis also suggests extreme nitrogen enrichment in the dense gas. Observations of young super star clusters in the Sunburst Arc have also revealed compact, high-density, nitrogen-enriched nebular gas, suggesting the local accumulation of massive-star ejecta \citep{Pascale2023,Pascale2024}. Together, these findings suggest that strong UV nitrogen emission can preferentially trace compact, dense, and chemically enriched gas rather than representing the chemical abundance of the entire galaxy. In this case, the nitrogen enhancement is not a galaxy-wide property, but instead confined to dense gas clouds surrounding WR stars and possibly LBVs in star-forming regions.}

Such localized chemical enrichment has been proposed in theoretical studies. Using three-dimensional radiation hydrodynamics simulations, \citet{Fukushima2024} show that compact, massive clouds ($\Sigma_\mathrm{cl}\gtrsim380\ M_\odot\ \mathrm{pc^{-2}}(M_\mathrm{cl}/10^8\ M_\odot)^{-1/5}$) can reach high N/O ratios via stellar winds from WR stars when the duration of star formation is longer than that of WR stars ($\sim3$ Myr) and shorter than that of SNe ($\sim10$ Myr). \red{Similarly, simulations of young super star cluster formation suggest that winds from VMSs can produce localized nitrogen enrichment in the surrounding dense gas \citep{Shi2026}.} Recent radiation hydrodynamic simulations further demonstrate that in dense gas clouds, efficient radiative cooling and turbulent mixing dissipate most of the wind energy, preventing efficient cloud disruption while allowing wind material to remain confined within the dense cloud \citep{Lancaster2021}. \red{Dense gas accumulated near young massive star clusters may also resist removal by subsequent CCSNe, as proposed for the compact nitrogen-enriched nebula surrounding the super star cluster Godzilla \citep{Pascale2024}.} This provides a physical mechanism by which nitrogen-rich stellar ejecta can accumulate locally instead of being rapidly mixed throughout the galaxy. A similar local enrichment scenario may also arise in dense CSM surrounding evolved massive stars, such as LBVs, whose nitrogen-rich ejecta can produce compact, high-density nebulae. In this local N-enrichment scenario, the broad N \textsc{iv}] component could trace the stellar wind or circumstellar ejecta, while the narrow N \textsc{iv}] emission originates from dense nitrogen-enriched gas confined in the immediate environment of the massive stars. The markedly higher electron density inferred from N \textsc{iv}] than from C \textsc{iii}] is qualitatively consistent with this interpretation. Although the present data do not uniquely distinguish between WR- and LBV-related enrichment, both scenarios \red{are consistent with high-ionization nitrogen lines originating from compact, locally enriched regions rather than the bulk H \textsc{ii} region. This localized enrichment scenario may also alleviate the difficulty of maintaining high N/O ratios against subsequent oxygen enrichment by CCSNe, as described in Section \ref{sec:introduction}. If the strong UV nitrogen emission predominantly originates from compact nitrogen-rich gas near massive stars, the surrounding diffuse H \textsc{ii} region need not exhibit similarly enhanced N/O ratios.} 
Consequently, abundance ratios derived from N \textsc{iv}] and N \textsc{iii}] may not represent the galaxy-wide chemical composition of GN-z11, but instead reflect the physical conditions and chemical enrichment of a localized dense region associated with massive-star ejecta. 

% Figure \ref{fig:structure} summarizes this proposed physical picture, illustrating how stellar winds from evolved massive stars may produce a locally nitrogen-enriched dense region embedded within a lower-density H \textsc{ii} region, thereby giving rise to the observed UV spectral features.

\red{
Figure \ref{fig:structure} illustrates the schematic view of the observed UV features of GN-z11. The bright ($M_\mathrm{UV}=-21.5$) and compact ($r_\mathrm{eff}=64$ pc) UV emission is likely dominated by young ($t_\mathrm{age}\simeq3$ Myr) massive stars, including O-type stars, WN stars, and LBVs, formed through intense star formation. These massive stars ionize the surrounding gas and drive stellar winds, as indicated by P-Cygni profiles of N \textsc{v}, Si \textsc{iv}, and C \textsc{iv}, as well as the broad N \textsc{iv}] (FWHM$=1640$ km s$^{-1}$) and tentative broad He \textsc{ii} (FWHM$=750$ km s$^{-1}$) emission. Stellar winds from these massive stars may locally enrich the  surrounding dense region ($n_\mathrm{e}\gtrsim10^{6.5}$ cm$^{-3}$) with nitrogen, giving rise to strong N \textsc{iv}] and possibly N \textsc{iii}] emission. In contrast, C \textsc{iii}], O \textsc{iii}], and Si \textsc{iii}] likely arise predominantly from the surrounding diffuse H \textsc{ii} region ($n_\mathrm{e}\lesssim10^{4}$ cm$^{-3}$). The extreme N/O inferred from the UV nitrogen lines may therefore preferentially trace this localized dense component rather than the galaxy-wide chemical abundance. On larger scales, stellar winds and/or the intense UV radiation field from massive stars may drive galactic scale outflows, as suggested by blue-shifted Si \textsc{iv} ($\Delta v=-438$ km s$^{-1}$) and C \textsc{iv} ($\Delta v=-464$ km s$^{-1}$) absorption. The red-shifted Ly$\alpha$ ($\Delta v=+453$ km s$^{-1}$) and C \textsc{iv} ($\Delta v=+394$ km s$^{-1}$) emission may further be affected by resonant scattering in the surrounding outflowing gas. Together, these results provide a unified physical picture in which young massive stars power the bright and compact UV emission of GN-z11, while their stellar winds and radiative feedback shape the surrounding gas. The photoionization modeling further supports a multi-component gas structure, with strong nitrogen emission arising predominantly from dense, highly ionized gas, which may be locally enriched in nitrogen by stellar winds, while galactic outflows redistribute gas on larger scales.
}

\section{Summary} 
\label{sec:summary}

In this study, we investigate the UV spectrum of a luminous compact galaxy, GN-z11 at $z=10.60$, using deep JWST/NIRSpec high-resolution IFU and medium-resolution MSA data obtained from the JADES, SPURS, and GO programs. After carefully reducing the IFU data by conducting $1/f$ noise corrections with \texttt{NSClean}, optimal extraction, and blank-sky-based error estimation, we obtain a 1D spectrum, consistent with the MSA spectra. Using \red{the complementary} MSA and IFU spectra, we explore the physical origin of the observed UV features of GN-z11. \red{We note that an independent analysis of the medium-resolution spectrum of GN-z11 by the SPURS team is presented in \citet{Chen2026b}.} We summarize our major findings below:

\begin{itemize}
    \item[1.] We identify prominent P-Cygni lines of N \textsc{v} $\lambda\lambda1238,1243$, Si \textsc{iv} $\lambda\lambda1394,1403$, and C \textsc{iv} $\lambda\lambda1548,1550$, together with a broad N \textsc{iv}] $\lambda\lambda1483,1486$ ($\mathrm{FWHM}=1640$ km s$^{-1}$) and a tentative broad He \textsc{ii} $\lambda1640$ ($\mathrm{FWHM}=750$ km s$^{-1}$) components. The P-Cygni profiles resemble those \red{observed in} local massive stars such as O-type stars and LBVs while the broad N \textsc{iv}] resembles that of WN stars, providing evidence for substantial stellar winds in GN-z11.

    \item[2.] \red{We quantitatively compare stellar and AGN interpretations by fitting the UV continuum, nebular emission, and ISM/CGM absorption simultaneously.} The stellar model is strongly favored over the AGN model based on statistical evaluation (\red{$\Delta\mathrm{WAIC}=-20.4$}) and naturally reproduces the P-Cygni profiles, particularly for N \textsc{v}, while the AGN model requires an unusually broad absorption component. The empirical comparison with local massive stars and the quantitative spectral fitting suggests that the UV continuum of GN-z11 is likely dominated by massive stars. 
    
    \item[3.] The stellar model fitting reveals the red-shifted nebular emission (Ly$\alpha$ and C \textsc{iv}) and blue-shifted ISM/CGM absorption (Si \textsc{iv} and C \textsc{iv}), which have comparable velocities of $\sim400-450$ km s$^{-1}$. This suggests the presence of galactic scale outflows, possibly driven by \red{feedback from the massive stellar population}.

    \item[4.] \red{The N \textsc{iv}], N \textsc{iii}], and C \textsc{iii}] multiplets, together with optical oxygen lines, reveal a strongly stratified density structure. While C \textsc{iii}] and oxygen lines traces relatively diffuse gas with $n_\mathrm{e}\lesssim10^{4}$ cm$^{-3}$, the nitrogen diagnostics extend to substantially higher densities, reaching $n_\mathrm{e}\gtrsim10^{6.5}$ cm$^{-3}$ for N \textsc{iv}]. This indicates that the strong nitrogen and carbon emission arises from physically distinct nebular components.} 

    \item[5.] \red{Our photoionization modeling of the UV-to-optical emission lines with \textsc{Cloudy} strongly favors a two-zone model over a one-zone model ($\Delta\mathrm{WAIC}=374.1$). The high-density zone has a higher ionization parameter and predominantly produces N \textsc{iv}] emission, while the low-density zone has a lower ionization parameter and dominates C \textsc{iii}] and [O \textsc{ii}] emission. A model with distinct N/O ratios between the two components is slightly favored over one with a common N/O ratio, although the difference is marginal ($\Delta\mathrm{WAIC}=1.3$).}

    \item[6.] \red{Our results suggest a unified picture in which young massive stars dominate the compact UV emission of GN-z11, while their nitrogen-enhanced stellar winds locally enrich the surrounding dense gas that produces the strong narrow nitrogen emission. The nitrogen enhancement inferred from the integrated spectrum may therefore reflect localized enrichment around massive stars rather than a galaxy-wide abundance pattern.}
    %These findings suggest that the apparent nitrogen enhancement estimated for GN-z11 may arise when strong narrow nitrogen emission originates from dense gas locally enriched in nitrogen by nitrogen-enhanced stellar winds and photoionized by nearby massive stars within the same star-forming  region.
\end{itemize}

%% Also note that the akcnowlodgment environment does not support long amounts of text. If you have a lot of people and institutions to acknowledge, do not use this command. Instead, create a new \section{Acknowledgments}.

\section*{Acknowledgements} 

\red{We thank the anonymous referee for the valuable comments that improved our manuscript.} We thank Tomokazu Kiyota for providing scripts of the IFU data reduction and for the helpful discussion on the IFU data.
\red{We thank Manuela Bischetti, John Chisholm, Liang Dai, Andrea Ferrara, Hajime Fukushima, Akio Ionue, Yuta Kageura, Roberto Maiolino, Charlotte Mason, Jorryt Matthee, Takashi Moriya, Pascal Oesch, Dan Stark, Stephania Salvadori, Yunjing Wu, and Hidenobu Yajima for the valuable discussions on this work.} We thank Yuki Isobe for providing the line-spread functions for NIRSpec.
This work is based on observations made with the NASA/ESA/CSA James Webb Space Telescope. The data were obtained from the Mikulski Archive for Space Telescopes (MAST) at the Space Telescope Science Institute (STScI), which is operated by the Association of Universities for Research in Astronomy (AURA), Inc., under NASA contract NAS 5-03127 for JWST. \red{The calibration level 3 data in MAST are available at \dataset[10.17909/d4b6-1z22]{https://doi.org/10.17909/d4b6-1z22}. In our analysis, we begin with the level 2 and 1 versions for the MSA and IFU data, respectively.} The JWST observations are associated with programs GO-9214 (SPURS), GTO-1181 (JADES), and GO-5086. We acknowledge the SPURS, JADES, and GO-5086 teams led by Charlotte Mason \& Dan Stark, Daniel Eisenstein \& Nora Lüetzgendorf, and Roberto Maiolino, respectively for developing their observation programs and for publicly releasing the data.
Based on observations obtained with the NASA/ESA Hubble Space Telescope, retrieved from the MAST at the STScI. STScI is operated by the Association of Universities for Research in Astronomy, Inc. under NASA contract NAS 5-26555.
This work benefited from discussions at the workshop ``Continuing the JWST Revolution: Understanding Early Galaxy Formation," hosted by the Munich Institute for Astro-, Particle-, and Bio-Physics (MIAPbP).
MN acknowledges support from  KAKENHI Grant Nos. 25KJ0828 through Japan Society for the Promotion of Science (JSPS).
MO acknowledges the supports from the World Premier International Research Center Initiative (WPI Initiative), MEXT, Japan, the joint research program of the Institute for Cosmic Ray Research (ICRR), the University of Tokyo, and KAKENHI (20H00180, 21H04467, 25H00674) through JSPS. The English in this paper was partially refined with the assistance of ChatGPT (OpenAI).


%% citation or url to the code inside ()s when available

\software{NumPy \citep{Harris2020}, matplotlib \citep{Hunter2007}, SciPy \citep{Virtanen2020}, Astropy \citep{Astropy2013,Astropy2018,Astropy2022}, \texttt{msaexp} \citep{Brammer2023}, \texttt{emcee} \citep{Foreman2013}, \texttt{STPSF} \citep{Perrin2025}}, BPASS v2.2.1 \citep{Eldridge2017,Stanway2018}, \textsc{Cloudy} v23.01 \citep{Ferland1998,Gunasekera2023}.





%\clearpage
%\bibliographystyle{aasjournal}
%\bibliographystyle{aasjournalv7}
\bibliography{library.bib}



\clearpage
\appendix
\restartappendixnumbering
\section{N \textsc{iii}] Electron-Density Diagnostics}
\label{Appendix:N3}

\begin{figure*}[ht!]
    \centering
    \includegraphics[width=0.99\linewidth]{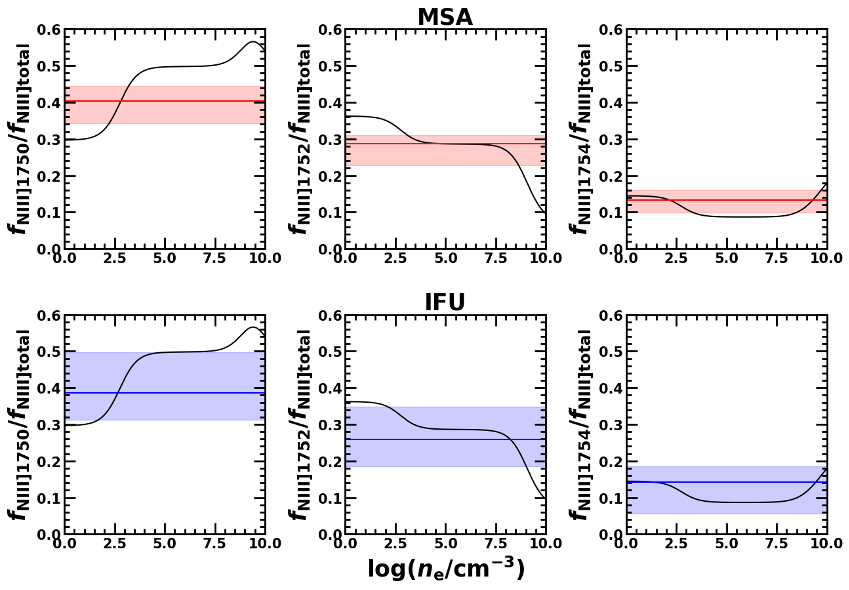}
    \caption{\red{Electron-density dependence of the N \textsc{iii}] line fractions. The top and bottom panels show the MSA and IFU results, respectively. The black curves represent the theoretical line fractions calculated with \texttt{PyNeb} as a function of electron density, where we assume $T_\mathrm{e}=14000$ K \citep{Alvarez-Marquez2025}. The red (blue) lines and shaded regions indicate the observed fractions derived from Gaussian fitting and their $1\sigma$ uncertainties for the MSA (IFU) spectra, respectively. The three columns correspond to the $\lambda1750$, $\lambda1752$, and $\lambda1754$ transitions, normalized by the total N \textsc{iii}] multiplet flux.}}
    \label{fig:n3_ratio}
\end{figure*}

\red{In Figure \ref{fig:n3_ratio}, we present the density dependence of the three independent N \textsc{iii}] line fractions, $\lambda1750$, $\lambda1752$, and $\lambda1754$, relative to the total multiplet flux. The theoretical predictions are calculated using \texttt{PyNeb}, while the observed fractions are derived from Gaussian fitting of the MSA and IFU spectra. The $\lambda1754$ fraction exhibits a non-monotonic dependence on electron density, potentially yielding degenerate solutions at low and high densities. Although the $\lambda1750$ and $\lambda1752$ fractions show less degeneracy, all three fractions are relatively insensitive to electron density over $n_\mathrm{e}\sim10^4$--$10^8\,\mathrm{cm}^{-3}$. Consequently, the N \textsc{iii}] multiplet provides limited density constraints over this range. These results highlight the importance of simultaneously modeling the full multiplet rather than relying on individual line fractions. We note that, although only the $\lambda1750$ fraction is presented in Figure \ref{fig:ne_measurement}, our electron-density measurements are obtained by forward modeling all five N \textsc{iii}] transitions simultaneously.}

\end{document}